\documentclass[12pt,a4paper]{article}
\pdfoutput=1
\usepackage{jheppub}
\usepackage{amsmath}
\usepackage{dsfont}
\usepackage{MnSymbol}
\usepackage{ulem}
\usepackage{natbib}
\usepackage[hang,flushmargin]{footmisc}

\makeatletter\renewcommand{\@biblabel}[1]{#1.}\makeatother

\def\be{\begin{equation}}
\def\ee{\end{equation}}
\def\ben{\begin{eqnarray}\displaystyle}
\def\een{\end{eqnarray}}

\def\nn{\nonumber}
\def\rd{{\rm d}}

\def\i{{\rm i}}
\def\e{{\rm e}}
\def\a{{\sf a}}
\def\b{{\sf b}}
\def\V{{\sf V}}

\def\S{{\sf S}}
\def\Q{{\sf Q}}
\def\P{{\sf P}}
\def\T{{\sf T}}

\def\ket#1{{ |#1\rangle}}
\def\bra#1{{ \langle#1|}}
\def\braket #1#2{{ \langle #1|#2 \rangle}}

\preprint{
{\small{\textsf{}}}}

\title{An elliptic Virasoro symmetry in 6d}

\author{Fabrizio Nieri}

\affiliation{Department of Physics and Astronomy, Uppsala University,\\
Box 516, SE-75120 Uppsala, Sweden.}

\emailAdd{fb.nieri@gmail.com}

\abstract{We define an elliptic deformation of the Virasoro algebra. We conjecture that the $\mathbb{R}^4\times \mathbb{T}^2$ Nekrasov partition function reproduces the chiral blocks of this algebra. We support this proposal by showing that at special points in the moduli space the 6d Nekrasov partition function reduces to the partition function of a 4d vortex theory supported on $\mathbb{R}^2\times \mathbb{T}^2$, which is in turn captured by a free field correlator of vertex operators and screening charges of the elliptic Virasoro algebra.}

\mathsubclass{81T20, 81T40, 81T60, 81R10, 81R50, 17B68.}


\begin{document}

\maketitle
\newpage

\section{Introduction}
The past few years have seen significant advances in our understanding of supersymmetric gauge theories. Such progress has been largely possible due to the developments of supersymmetric localization techniques which have allowed many exact results to be obtained. One of these is the discovery by Alday, Gaiotto and Tachikawa (AGT) \cite{Alday:2009aq} that certain BPS observables of class $\mathcal{S}$ theories \cite{Gaiotto:2009we} of $A_1$ type can be computed in Liouville CFT, or Toda for higher rank \cite{Wyllard:2009hg}. In particular, the AGT correspondence identifies the $\mathbb{R}^4$ Nekrasov instanton partition function ($\mathcal{Z}^{\mathbb{R}^4}_{\rm inst}$) \cite{Nekrasov:2002qd,Nekrasov:2003rj} with the chiral blocks \cite{Belavin:1984vu} of the Virasoro algebra (or $W_{M}$ algebra for $A_{M-1}$ theories)
\be
\mathcal{Z}^{\mathbb{R}^4}_{\rm inst}\simeq\bra {\gamma_\infty} \prod_{i=1}^N V_{\gamma_i}(x_i)\ket{\gamma_0}_{\rm Vir}~.\nn
\ee
The AGT relation is a powerful tool to get further insights into the gauge dynamics as certain aspects can be efficiently addressed in the 2d CFT side. For example, the study of defect operators \cite{Alday:2009fs,Drukker:2009id,Frenkel:2015rda,Coman:2015lna,Gomis:2014eya,Hosomichi:2010vh,Teschner:2012em,Doroud:2012xw} (for a recent review we refer to \cite{Teschner:2014oja,Okuda:2014fja,Gukov:2014gja}). 

It seems to be quite important to understand whether AGT-like relations exist in other dimensions as well, in particular in 6d where much of the 4d physics finds its natural origin. One of the main motivations behind this work was indeed to explore the possibility of studying 6d theories through AGT inspired methods, a topic which was previously addressed also in \cite{Nekrasov:2012xe,Nekrasov:2013xda}. Our results can be summarized in the proposal
\be
\mathcal{Z}^{\mathbb{R}^4\times \mathbb{T}^2}_{\rm inst}\simeq\bra {\gamma_\infty} \prod_{i=1}^N V_{\gamma_i}(x_i)\ket{\gamma_0}_{e\textrm{-Vir}}~,\nn
\ee
where the l.h.s. captures the supersymmetric partition function of a 6d $(1,0)$ theory on a torus which can be engineered in M-theory by two M5-branes probing a transverse $A_{N-1}$ singularity \cite{Blum:1997mm}, while the r.h.s. represents the chiral blocks of an elliptically deformed Virasoro algebra, which we define in this paper.

This result can be read as the natural 1-parameter deformation (table \ref{tab1}) 
\begin{table}[!ht]
\begin{center}
\begin{tabular}{c|c}
$\mathcal{Z}_{\rm inst}$ on~&~Chiral blocks of \\
\hline
$\mathbb{R}^4$~&~$\phantom{q\;}$Virasoro\\
$\mathbb{R}^4\times S^1$~&~$q$-Virasoro\\
$\mathbb{R}^4\times \mathbb{T}^2$~&~$e$-Virasoro
\end{tabular}
\caption{The AGT relation in various dimensions.}\label{tab1}
\end{center}
\end{table}
of the 5d AGT relation \cite{Awata:2009ur}, which we briefly recall here to pave the way for our analysis in this work. The $\mathbb{R}^4$ Nekrasov  partition function and the Virasoro algebra have both a natural trigonometric deformation. The deformation of the former corresponds to the $\mathbb{R}^4\times S^1$ Nekrasov partition function \cite{Nekrasov:2002qd,Nekrasov:2003rj}, while the deformation of the latter corresponds to the $q$-Virasoro algebra ($q$-$W_M$ algebra) of \cite{Shiraishi:1995rp,Awata:1995zk,Feigin:1995sf}. The identification of the two deformations predicts the 5d AGT correspondence
\be
\mathcal{Z}^{\mathbb{R}^4\times S^1}_{\rm inst}\simeq\bra {\gamma_\infty} \prod_{i=1}^N V_{\gamma_i}(x_i)\ket{\gamma_0}_{q\textrm{-Vir}}~.\nn
\ee
Evidences supporting this idea were extensively discussed for example in \cite{Awata:2010yy,Mironov:2011dk,Carlsson:2013jka}, and more recently in \cite{Zenkevich:2014lca,Morozov:2015xya}. A complete 5d AGT relation beyond the chiral level was also proposed in \cite{Nieri:2013yra,Nieri:2013vba}, where the $S^5$ \cite{Hosomichi:2012ek,Kallen:2012cs,Kallen:2012va,Imamura:2012bm,Lockhart:2012vp,Kim:2012qf,Kim:2012ava,Minahan:2013jwa} and $S^4\times S^1$ \cite{Kim:2012gu,Terashima:2012ra,Iqbal:2012xm} partition functions of the 5d lift of class $\mathcal{S}$ theories \cite{Benini:2009gi} of $A_1$ type were shown to be described by correlators in two distinct QFTs with $q$-Virasoro symmetry, and hence called $q$-CFTs (see also \cite{Kozcaz:2010af,Mitev:2014isa,Isachenkov:2014eya} for an analysis of the higher rank case).

Another neat argument in favor of the 5d AGT correspondence, which we also adopt in this paper for the 6d analysis, was  given in \cite{Aganagic:2013tta} (see also the review \cite{Aganagic:2014kja}). It was shown that the $\mathbb{R}^4\times S^1$ Nekrasov instanton partition function of the $U(N)$ theory with $N$ fundamental and anti-fundamental flavors reproduces, upon suitable specializations $a=a_*(r)$ of the Coulomb branch moduli, the $(N+2)$-point chiral blocks of the $q$-Virasoro algebra in the Dotsenko-Fateev free field integral representation \cite{Awata:2010yy,Lukyanov:1994re}
\be
\mathcal{Z}^{\mathbb{R}^4\times S^1}_{{\rm inst}}\Big|_{a=a_*(r)}\simeq \oint\!\!\rd z\; \bra{\gamma_{\infty}}\prod_{i=1}^N V_{\gamma_i}(x_i)\prod_{j=1}^r S(z_j)\ket{\gamma_0}_{q\textrm{-Vir}}~,\nn
\ee
where the operator $S(z)$ denotes the screening current and the integral is computed by residues for specific choices of integration contours. The specialization of the Coulomb branch moduli corresponds to the root of the Higgs branch where vortex solutions exist and the dynamics can be effectively described by a 1/2 BPS codimension 2 theory on $\mathbb{R}^2_{\epsilon}\times S^1$. Partition functions of 3d $\mathcal{N}=2$  gauge theories compactified on $\mathbb{R}^2_{\epsilon}\times S^1$ can be computed by means of the 3d holomorphic block integrals ($\mathcal{B}^{\rm 3d}$) introduced in \cite{Beem:2012mb} (see also \cite{Pasquetti:2011fj} for previous work on 3d block factorization and \cite{Yoshida:2014ssa} for a derivation of block integrals through localization)
\be
\mathcal{B}^{\rm 3d}=\oint\!\!\frac{\rd z}{2\pi\i z}\Upsilon^{\rm 3d}(z)~,\nn
\ee
where the integral kernel $\Upsilon^{\rm 3d}(z)$ is a meromorphic function determined by the specific theory and the integration is over a basis of middle dimensional cycles in $(\mathbb{C}^\times)^{|G|}$, $G$ being the gauge group. It was pointed out in \cite{Aganagic:2013tta} that the free field integral representation of $q$-Virasoro chiral blocks manifestly matches the 3d block integrals of the $U(r)$ theory with $N$ fundamental and anti-fundamental flavors, 1 adjoint and Fayet-Iliopoulos term. Combining all these observations, one gets the  identifications
\be
\mathcal{Z}^{\mathbb{R}^4\times S^1}_{{\rm inst}}\Big|_{a=a_*(r)}\simeq \oint\!\!\rd z\; \bra{\gamma_{\infty}}\prod_{i=1}^N V_{\gamma_i}(x_i)\prod_{j=1}^r S(z_j)\ket{\gamma_0}_{q\textrm{-Vir}}\simeq\mathcal{B}^{\rm 3d}[U(r)]~.\nn
\ee
In particular, the two parameters $q,t$ of the $q$-Virasoro algebra are identified with the $\epsilon_{1,2}$ parameters of the 5d $\Omega$-background $\mathbb{R}^4_{\epsilon_1,\epsilon_2}\times S^1$, and with the angular momentum fugacity $\epsilon$ and adjoint real mass in the 3d theory. The choice $a=a_*(r)$ determines the rank $r$ of the 3d gauge group and the number of screening currents, which must be in turn distributed amongst the $N$ flavors and insertion points according to a choice of partition $r=\sum_{a=1}^N r_a$. This choice corresponds to an integration contour and provides additional discrete variables (filling fractions \cite{Dijkgraaf:2009pc,Dijkgraaf:2002vw,Dijkgraaf:2002fc}) entering the allowed values of the internal momenta in the correlator. This kind of ``triality" can be extended to quiver gauge theories and $q$-$W_M$ correlators \cite{Aganagic:2014oia} (the generalization to $DE$ root systems and applications to little string theories \cite{Seiberg:1997zk,Berkooz:1997cq,Losev:1997hx} can be found in \cite{Aganagic:2015cta}).

The natural lift of the 3d setup is provided by 4d $\mathcal{N}=1$ gauge theories compactified on $\mathbb{R}^2_{\epsilon}\times \mathbb{T}^2$. Partition functions on this background can be computed through the 4d holomorphic block integrals ($\mathcal{B}^{\rm 4d}$) introduced in \cite{Nieri:2015yia}. These objects can be thought of as a 1-parameter deformation of 3d block integrals, in the sense that
\be
\mathcal{B}^{\rm 4d}=\oint\!\!\frac{\rd z}{2\pi\i z}\Upsilon^{\rm 4d}(z)
~~\xrightarrow{\textrm{3d limit}}~~\mathcal{B}^{\rm 3d}=\oint\!\!\frac{\rd z}{2\pi\i z}\Upsilon^{\rm 3d}(z)\nn~.
\ee
Due to the algebraic interpretation of 3d holomorphic blocks with adjoint matter as $q$-Virasoro chiral blocks in free field representation, we are naturally led to ask whether 4d holomorphic blocks with adjoint matter can be similarly interpreted as chiral blocks of an elliptic deformation of the Virasoro algebra. A central result of the current paper is that we can give an affirmative answer to that question, namely
\be
\mathcal{B}^{\rm 4d}[U(r)]\simeq\oint\!\!\rd z\; \bra{\gamma_{\infty}}\prod_{i=1}^N V_{\gamma_i}(x_i)\prod_{j=1}^r S(z_j)\ket{\gamma_0}_{e\textrm{-Vir}}~.\nn
\ee 
In turn, we are also able to match the evaluation of the 4d block integral with the topological string partition function on the Calabi-Yau geometry obtained by gluing two periodic strips \cite{Hollowood:2003cv,Haghighat:2013tka,Hohenegger:2013ala}, for special values of the K\"ahler moduli. The latter captures the M-theory partition function of two M5-branes extending in $\mathbb{R}^4\times\mathbb{T}^2$ and probing a transverse $A_{N-1}$ singularity, and hence we can also argue that the 4d gauge theories under examination describe vortices of 6d $(1,0)$ theories so engineered. This chain of results, together with large $r$ duality \cite{Gopakumar:1998ki,Cachazo:2001jy,Aganagic:2002wv,Aganagic:2011mi}, strongly supports a 6d version of the AGT correspondence (discussed also in \cite{Tan:2013xba} from the M-theory perspective) by identifying generic chiral blocks of the elliptic Virasoro algebra with $\mathbb{R}^4\times\mathbb{T}^2$ Nekrasov instanton partition functions.  Lately, 6d $(1,0)$ theories have attracted much attention \cite{Heckman:2015bfa,Heckman:2013pva,DelZotto:2015rca,Bhardwaj:2015xxa,Hohenegger:2015cba,Gadde:2015tra,Haghighat:2015coa,Ohmori:2015pua,Ohmori:2015pia,Zafrir:2015rga,Kim:2015fxa,Ohmori:2015tka}. We hope that the methods developed in this work will be useful to get further insights into the elusive 6d physics. 

The rest of this paper is organized as follows.  In section \ref{seceva}, we define an elliptic deformation of the Virasoro algebra and compute free field correlators. In section \ref{4dholblocks}, we briefly review the 4d block integral formalism and we consider its application to the $U(r)$ theory with adjoint matter, manifestly matching free field correlators of the elliptic Virasoro algebra. In section \ref{sec6dnek}, we compute the $\mathbb{R}^4\times \mathbb{T}^2$ Nekrasov instanton partition function by using the topological vertex on the periodic strip, and we show that at specific points it coincides with the elliptic vortex partition function of the 4d theory. In section \ref{out}, we discuss our results further as well as interesting directions for future research. Other and more technical aspects of this work are discussed in several appendices.
\bigskip


\section{Elliptic Virasoro Algebra}\label{seceva}
In this section we define a 1-parameter deformation of the Deformed Virasoro Algebra (DVA or $q$-Virasoro) of \cite{Shiraishi:1995rp}, which we call the Elliptic Virasoro Algebra (EVA or $e$-Virasoro). We give a free field representation of the EVA and find its screening currents. We then compute free field correlators of suitably defined vertex operators and screening charges. The special functions used in this section are collected in appendix \ref{appell}, while part of our notation is set in appendix \ref{freeboson}.

\subsection{Defining relation}
We define the EVA to an be associative algebra generated by the coefficients of the current $T(z)=\sum_{n\in\mathbb{Z}}T_n z^{-n}$, with the defining relations encoded by
\be\label{EVAalgebra}
f\left(\frac{w}{z}\right)T(z)T(w)-T(w)T(z)f\left(\frac{z}{w}\right)=-\frac{\Theta(q;q')\Theta(t^{-1};q')}{(q';q')_\infty^2\Theta(p;q')}\left(\delta\left(p\frac{w}{z}\right)-\delta\left(p^{-1}\frac{w}{z}\right)\right)~,
\ee
where the coefficients of the structure function $f(x)=\sum_{\ell\in\mathbb{Z}}f_\ell x^\ell$ are defined by the series expansion of
\be
f(x)=\frac{\Gamma(x;p^2,q')\Gamma(p^2q^{-1}x;p^2,q')\Gamma(p q x;p^2,q')}{\Gamma(p^2x;p^2,q')\Gamma(p q^{-1}x;p^2,q')\Gamma(q x;p^2,q')}~,
\ee
with the elliptic Gamma function defined in (\ref{ellgamma}) in the region $|p^2|,|q'|<1$, 
while $\delta(x)=\sum_{n\in\mathbb{Z}}x^n$ is the multiplicative $\delta$ function, i.e. $\delta(x)\phi(x)=\delta(x)\phi(1)$ for any Laurent series $\phi(x)$. The parameters $q,t,q'$ are complex, $p=qt^{-1}$, and for later convenience we also define $\beta\in\mathbb{C}$ such that $t=q^\beta$. 

{\it Remark}. The associativity of the algebra is equivalent to the requirement \cite{Awata:1996fq,Odake:1999un}
\be\label{YB}
f(x)f(x p^{-1})-f(x^{-1})f(p x^{-1})= \kappa( \delta(x p^{-1})-\delta(x))~,
\ee
for some constant coefficient $\kappa$, arising from the Yang-Baxter equation for $T(z)$. The validity of (\ref{YB}) can be explicitly verified by using (\ref{fgamma}), (\ref{gamma1}), (\ref{gamma2}). 

The choice of the structure function is motivated by the property above, the construction given in appendix \ref{ding}, and by the fact that in the limit $q'\to 0$, (\ref{EVAalgebra}) manifestly reduces to the defining relation of the DVA by using 
\be
\Gamma(x;p^2,0)=\frac{1}{(x;p^2)_\infty}~,\quad \Theta(x;0)=1-x~.
\ee
This trigonometric limit can be verified at various stages of our construction below. 

{\it Remark}. By comparing the coefficients of $z^{-n}w^{-m}$, the defining relation (\ref{EVAalgebra})  is equivalent to the quadratic relation
\be
\sum_{\ell\in\mathbb{Z}}f_\ell (T_{n-\ell}T_{m+\ell}-T_{m-\ell}T_{n+\ell})=-\frac{\Theta(q;q')\Theta(t^{-1};q')}{(q';q')_\infty^2\Theta(p;q')}(p^n-p^{-n})\delta_{n+m,0}~.
\ee

\subsection{Free field representation}
In order to find a free field representation of the EVA, we introduce two commuting families of quantum bosonic oscillators $\{\alpha_n,\beta_n,n\in\mathbb{Z}\backslash\{0\}\}$. They satisfy the commutation relations (we display the non-trivial relations only) 
\be\label{roottype}
\begin{split}
[\alpha_n,\alpha_m]&=\frac{1}{n}(1-q'^{|n|})(q^{\frac{n}{2}}-q^{-\frac{n}{2}})(t^{\frac{n}{2}}-t^{-\frac{n}{2}})(p^{\frac{n}{2}}+p^{-\frac{n}{2}})\delta_{m+n,0}~,\\
[\beta_n,\beta_m]&=\frac{q'^{|n|}}{n}(1-q'^{|n|})(q^{\frac{n}{2}}-q^{-\frac{n}{2}})(t^{\frac{n}{2}}-t^{-\frac{n}{2}})(p^{\frac{n}{2}}+p^{-\frac{n}{2}})\delta_{m+n,0}~.
\end{split}
\ee
We also introduce zero mode operators $\P,\Q$ commuting with all the oscillators and normalized according to
\be
[\P,\Q]=2~.
\ee

We then define the currents
\be
\Lambda_{\sigma}(z)=\; :\e^{\sigma\sum_{n\neq 0}\frac{z^{-n}}{(1+p^{-\sigma n})(1-q'^{|n|})}\alpha_n}\e^{\sigma\sum_{n\neq 0}\frac{z^{n}}{(1+p^{\sigma n})(1-q'^{|n|})}\beta_n}: q^{\sigma\frac{\sqrt{\beta}}{2}\P}p^{\sigma\frac{1}{2}}~,\quad \sigma\in\{+,-\}~.
\ee
The symbol $:~:$ denotes normal ordering, i.e. all the positive oscillators are placed to the right of the negative ones, and $\P$ to the right of the $\Q$. Using the commutation relations (\ref{roottype}), the definition (\ref{ellgamma}) of the elliptic Gamma function and the free boson tools summarized in appendix \ref{freeboson}, we can verify that the current
\be\label{EVAff}
\T(z)=\Lambda_{+}(z)+\Lambda_{-}(z)=\sum_{n\in\mathbb{Z}}\T_n z^{-n}~
\ee
satisfies the defining relation (\ref{EVAalgebra}) of the EVA. The explicit verification of this claim is straightforward but lengthy, and hence presented in appendix \ref{appfreeEVA}. The key relations to be used are 
\be\label{OPELambda}
\Lambda_\sigma(z)\Lambda_\rho(w)=\; :\Lambda_\sigma(z)\Lambda_\rho(w):f_{\sigma,\rho}(w z^{-1})^{-1}~,\quad (\sigma,\rho)\in\{\pm,\pm\}~,
\ee
where
\be\label{ffunction}
f_{\sigma,\rho}(x)=\left(\frac{\Gamma(p^{\frac{\sigma\cdot 1-\rho\cdot 1}{2}}x;p^2,q')\Gamma(p^{\frac{\sigma\cdot 1-\rho\cdot 1}{2}}p^2q^{-1}x;p^2,q')\Gamma(p^{\frac{\sigma\cdot 1-\rho\cdot 1}{2}}p q x;p^2,q')}{\Gamma(p^{\frac{\sigma\cdot 1-\rho\cdot 1}{2}}p^2x;p^2,q')\Gamma(p^{\frac{\sigma\cdot 1-\rho\cdot 1}{2}}p q^{-1}x;p^2,q')\Gamma(p^{\frac{\sigma\cdot 1-\rho\cdot 1}{2}}q x;p^2,q')}\right)^{\sigma\rho\;\cdot 1}~,
\ee
and
\be\label{LL1}
:\Lambda_+(z)\Lambda_-(p^{-1}z):\;=\mathds{1}~.
\ee
One also needs
\be\label{fgamma}
f(x)=f_{+,+}(x)=f_{-,-}(x)=f_{+,-}(x)\gamma(p^{\frac{1}{2}}x)=f_{-,+}(x)\gamma(p^{-\frac{1}{2}}x)~,
\ee
where
\be\label{gamma1}
\gamma(x)=\frac{\Theta(p^{\frac{1}{2}}q^{-1}x;q')\Theta(p^{-\frac{1}{2}}q x;q')}{\Theta(p^{\frac{1}{2}}x;q')\Theta(p^{-\frac{1}{2}}x;q')}~,
\ee
as well as the equality
\be\label{gamma2}
\gamma(x)-\gamma(x^{-1})=-\frac{\Theta(q;q')\Theta(t^{-1};q')}{(q';q')_\infty^2\Theta(p;q')}\left(\delta(p^{\frac{1}{2}}x)-\delta(p^{-\frac{1}{2}}x)\right)~,
\ee
which follows from the representation (\ref{delta}) of the $\delta$ function. More details along with technical comments are given  in appendix \ref{appfreeEVA}.

{\it Remark}. Using $(M-1)$-dimensional oscillators $\vec \alpha_n$, $\vec \beta_n$ associated to the $A_{M-1}$ root system, it is possible to extend our construction to define an elliptic version of the $W_{M}$ algebra, along the lines of \cite{Awata:1995zk,Feigin:1995sf,frenkelresh:1997}.

\subsection{Screening currents}
The screening current $\S(z)$ of the EVA in the free field representation (\ref{EVAff}) is defined by the relation
\be\label{TSrel}
[\T_n,\S(w)]=\frac{\rd }{\rd_q w}{\sf A}_n(w)=\frac{{\sf A}_n(q^{\frac{1}{2}}w)-{\sf A}_n(q^{-\frac{1}{2}}w)}{w(q^{\frac{1}{2}}-q^{-\frac{1}{2}})}
\ee
for some operator ${\sf A}_n(w)$, so that the EVA generators and the screening charge $\oint\!\! \rd w\;\S(w)$ commute for a suitable $q$-invariant integration contour (e.g. around the origin). With the definition
\be\label{screening}
\S(z)=\, :\e^{-\sum_{n\neq 0}\frac{z^{-n}}{(q^{n/2}-q^{-n/2})(1-q'^{|n|})}\alpha_n}\e^{\sum_{n\neq 0}\frac{z^{n}}{(q^{n/2}-q^{-n/2})(1-q'^{|n|})}\beta_n}:\e^{\sqrt{\beta}\Q}z^{\sqrt{\beta}\P}~,
\ee
we can verify that (\ref{TSrel}) is satisfied. Since this particular choice of screening current will be crucial for actual computations, in appendix \ref{appS} we explicitly show how the claim (\ref{TSrel}) can be verified. 

{\it Remark}. Another screening current can also be defined through the map $q\to t$, $t\to q$, $\alpha_n\to-\alpha_n$, $\beta_n\to -\beta_n$, $\sqrt{\beta}\to -1/\sqrt{\beta}$, but we do not need it for our purposes. 

The product of several screening currents can be written as
\be\label{Sprod}
\prod_{i=1}^r\S(z_i)
=\; :\prod_{i=1}^r\S(z_i): \times \prod_{1\leq i\neq j\leq r}\frac{\Gamma(t z_j z_i^{-1};q,q')}{\Gamma(z_j z_i^{-1};q,q')}\prod_{1\leq i<j\leq r}\frac{\Theta(t z_i z_j^{-1};q)}{\Theta(z_i z_j^{-1};q)}\times\prod_{i=1}^{r}z_i^{2\beta(r-i)} ~.
\ee
The last factor arises from the normal ordering of the zero modes
\be 
\prod_{i}\left(\e^{\sqrt{\beta}\Q}z_i^{\sqrt{\beta}\P}\right)=
\prod_{i}\e^{\sqrt{\beta}\Q}\prod_{i}z_i^{\sqrt{\beta}\P}\times \prod_{i<j}z_i^{2 \beta}~,
\ee
and it can also be rewritten as
\be\label{class}
\prod_{i<j}z_i^{2 \beta}=\prod_{i=1}^r z_i^{2\beta(r-i)}=\prod_{i=1}^r\frac{z_i^{\sqrt{\beta}(\sqrt{\beta}r-Q)}}{z_i}\times \prod_{1\leq i<j\leq r}\left(\frac{z_i}{z_j}\right)^\beta~,
\ee
where we defined
\be
Q=\sqrt{\beta}-\frac{1}{\sqrt{\beta}}~.
\ee
The last factor in the r.h.s. of (\ref{class}) can be put with the $\Theta$'s in (\ref{Sprod}) to form the $q$-constant\footnote{This is a function invariant w.r.t. to $q$-shifts of its arguments, i.e. $c_\beta(z;q)|_{z_i\to q z_i}=c_\beta(z;q)$.}
\be\label{cq}
c_\beta(z;q)=\prod_{1\leq i<j\leq r}\left(\frac{z_i}{z_j}\right)^\beta\frac{\Theta(t z_i z_j^{-1};q)}{\Theta(z_i z_j^{-1};q)}~,
\ee
while we can use the product of the $\Gamma$'s in (\ref{Sprod}) to define an elliptic Vandermonde-like determinant 
\be\label{DE}
\Delta_E(z)=\prod_{1\leq i\neq j\leq r}\frac{\Gamma(t z_i z_j^{-1};q,q')}{\Gamma(z_i z_j^{-1};q,q')}~.
\ee
When considering free field correlators with integrated screening currents, this object will provide the integration measure. In fact, we can simply forget about the $q$-constant (\ref{cq}) because of the integration contour that we will prescribe (see discussion in section \ref{4dholblocks}).

\subsection{Vertex operators}
Let us define the following vertex operator built out of the bosonic oscillators and zero modes
\be
\V_u(x)=\; :\e^{\sum_{n\neq 0}\frac{[u]_n x^{-n}}{(q^{n/2}-q^{-n/2})(1-q'^{|n|})}\alpha_n} \e^{-\sum_{n\neq 0}\frac{[u]_n x^{n}}{(q^{n/2}-q^{-n/2})(1-q'^{|n|})}\beta_n}:\e^{-\frac{\gamma}{2}\sqrt{\beta}\Q}x^{-\frac{\gamma}{2}\sqrt{\beta}\P}~,
\ee
where
\be
[u]_n=\frac{u^{\frac{n}{2}}-u^{-\frac{n}{2}}}{(t^{\frac{n}{2}}-t^{-\frac{n}{2}})(p^{\frac{n}{2}}+p^{-\frac{n}{2}})}~,\quad u=t^\gamma~.
\ee
The momentum $\gamma$ (or equivalently $u$) is a free parameter labelling the vertex operator. The ``OPE" between this vertex operator and the screening current can be written as
\be\label{VSprod}
\V_u(x)\S(z)
=\; :\V_u(x)\S(z):\times \frac{\Gamma(q^{\frac{1}{2}}u^{-\frac{1}{2}}z x^{-1};q,q')}{\Gamma(q^{\frac{1}{2}}u^{\frac{1}{2}}z x^{-1};q,q')}\times  x^{-\beta\gamma}~,
\ee
where the last factor arises from the normal ordering of the zero modes
\be 
x^{-\frac{\gamma}{2}\sqrt{\beta}\P}\e^{\sqrt{\beta}\Q}=
\e^{\sqrt{\beta}\Q} x^{-\frac{\gamma}{2}\sqrt{\beta}\P}\times x^{-\beta\gamma}~.
\ee
In the following, we do not need the explicit form of the ``OPE" between vertex operators alone.

\subsection{Correlators}\label{seccorr}
Let us start by defining the zero momentum Fock space  $\mathcal{F}_0$. It is the left module over the oscillator algebra (\ref{roottype}) generated by the vacuum $\ket{0}$ defined by 
\be
\alpha_{n}\ket{0}=\beta_{n}\ket{0}=0~,\quad n\in\mathbb{Z}_{>0}~,
\ee
namely
\be
\mathcal{F}_0={\rm Span}\Big\{\alpha_{-\mu}\beta_{-\nu}\ket{0}~,\mu,\nu\in\mathcal{P}\Big\}~,
\ee 
where $\mathcal{P}$ is the set of partitions, and for length $\ell(\mu)$, $\ell(\nu)$ partitions we defined $\alpha_{-\mu}=\alpha_{-\mu_1}\;\cdots\;\alpha_{-\mu_{\ell(\mu)}}$, $\beta_{-\nu}=\beta_{-\nu_1}\;\cdots\;\beta_{-\nu_{\ell(\nu)}}$. Let us also define the dual (right) Fock module 
\be
\mathcal{F}_0^*={\rm Span}\Big\{\bra{0}\alpha_{\mu}\beta_{\nu}~,\mu,\nu\in\mathcal{P}\Big\}~,
\ee 
generated by the dual vacuum $\bra{0}$ defined by $\bra{0}\alpha_{-n}\beta_{-m}=0$, $n,m\in\mathbb{Z}_{>0}$.\footnote{We identify $\alpha_n^\dagger=\alpha_{-n}$, $\beta_n^\dagger=\beta_{-n}$.} Acting with the exponential of the zero mode operator $\Q$ on the neutral vacua, we can define charged Fock vacua generating the charged Fock modules $\mathcal{F}_\gamma$, $\mathcal{F}^*_\gamma$
\be
\ket{\gamma}=\e^{\frac{\gamma}{2} {\Q}}\ket{0}~,\quad \bra{\gamma}=\bra{0}\e^{-\frac{\gamma}{2}{\Q}}~,\quad \gamma\in\mathbb{C}~.
\ee
The charged vacua are eigenstates of the momentum operator $\P$
\be
\P\ket{\gamma}=\gamma\ket{\gamma}~,\quad \bra{\gamma}\P=\gamma\bra{\gamma} ~,
\ee
and we define the following pairing between the left and right charged Fock modules
\be
\braket{\gamma}{\gamma'}=\delta_{\gamma,\gamma'}~.
\ee

We are now ready to compute correlators of $N$ vertex operators and $r$ integrated screening currents between external Fock states. We start by using the manipulations of the previous subsection to write
\begin{multline}\label{VScorr}
\prod_{i=1}^N\V_{u_i}(x_i)\prod_{i=1}^r\S(z_i)=\; :\prod_{i=1}^N\V_{u_i}(x_i)\prod_{i=1}^r\S(z_i):\times {\rm ``OPE"}\times \prod_{j=1}^N  x_j^{-\beta r\gamma_j}\times\\
\times c_\beta(z;q)\times \Delta_E(z)\prod_{i=1}^r\frac{z_i^{\sqrt{\beta}(\sqrt{\beta}r-Q)}}{z_i}\prod_{i=1}^r\prod_{j=1}^N \frac{\Gamma(q^{\frac{1}{2}}u_j^{-\frac{1}{2}}z_i x_j^{-1};q,q')}{\Gamma(q^{\frac{1}{2}}u_j^{\frac{1}{2}}z_i x_j^{-1};q,q')}~,
\end{multline}
where we set $u_j=t^{\gamma_j}$. The ${\rm ``OPE"}$ prefactor denotes all the normal ordering terms arising from vertex operators alone, which are not important for the present analysis and hence will be neglected in the following. Sandwiching (\ref{VScorr}) between two Fock states $\ket{\gamma_0}$ and $\bra{\gamma_\infty}$ we get (up to constant, i.e. $z_i$-independent factors)
\begin{multline}
\bra{\gamma_\infty}\prod_{i=1}^N \V_{u_i}(x_i)\prod_{i=1}^r\S(z_i)\ket{\gamma_0}\propto\braket{\gamma_\infty}{\gamma_0+\sqrt{\beta}(2r-\sum_i\gamma_i)}\times\\
\times c_\beta(z;q)\times \Delta_E(z)\prod_{i=1}^r\frac{z_i^{\sqrt{\beta}(\gamma_0+\sqrt{\beta}r-Q)}}{z_i} \prod_{i=1}^r\prod_{j=1}^N\frac{\Gamma(q^{\frac{1}{2}}u_j^{-\frac{1}{2}}z_i x_j^{-1};q,q')}{\Gamma(q^{\frac{1}{2}}u_j^{\frac{1}{2}}z_i x_j^{-1};q,q')}~,
\end{multline}
which is non-zero provided the neutrality condition $\sqrt{\beta}(2r-\sum_{i=1}^N \gamma_i)+\gamma_0-\gamma_\infty=0$ holds. In this case, the correlator integrated over the positions of the screening currents reads
\begin{align}
G^{(N)}_{\gamma_\infty,\gamma_0}&=\oint\prod_{i=1}^r\frac{\rd z_i}{2\pi \i}\bra{\gamma_\infty}\prod_{i=1}^N \V_{u_i}(x_i)\prod_{i=1}^r\S(z_i)\ket{\gamma_0}\propto\nn\\
&\propto \oint\prod_{i=1}^r\frac{\rd z_i}{2\pi \i z_i}\;c_\beta(z;q)\;\Delta_E(z)\prod_{i=1}^r z_i^{\sqrt{\beta}(\gamma_0+\sqrt{\beta}r-Q)}\prod_{i=1}^r\prod_{j=1}^N \frac{\Gamma(q^{\frac{1}{2}}u_j^{-\frac{1}{2}}z_i x_j^{-1};q,q')}{\Gamma(q^{\frac{1}{2}}u_j^{\frac{1}{2}}z_i x_j^{-1};q,q')}~,
\end{align}
where we left the integration contour momentarily unspecified. For later purposes, it is useful to reorder the integration variables $z_i\to z_{r-i+1}$, perform the change $z_i\to z_i^{-1}$, and set $q^{1/2}u_j^{-1/2}x_j^{-1}=y_j$. The combination of these transformations acts as the identity on $\Delta_E(z)$ and $c_\beta(z;q)$, and then we have
\be\label{Npoint}
G^{(N)}_{\gamma_\infty,\gamma_0}\propto \oint\prod_{i=1}^r\frac{\rd z_i}{2\pi \i z_i}\;c_\beta(z;q)\;\Delta_E(z)\prod_{i=1}^r z_i^{\sqrt{\beta}(Q-\gamma_0-\sqrt{\beta}r)}\prod_{i=1}^r\prod_{j=1}^N\frac{\Gamma(y_j z_i^{-1};q,q')}{\Gamma(u_j y_j z_i^{-1};q,q')}~.
\ee
This integral looks like an elliptic deformation of the Dotsenko-Fateev representation of the chiral blocks of the DVA.\footnote{For the role of Dotsenko-Fateev integrals in the 5d AGT duality we refer to \cite{Awata:2010yy,Mironov:2011dk,Zenkevich:2014lca,Morozov:2015xya,Aganagic:2013tta,Aganagic:2014oia}.} In fact, in the trigonometric limit $q'\to 0$ we get
\be
\lim_{q'\to 0}G^{(N)}_{\gamma_\infty,\gamma_0}\propto \oint\prod_{i=1}^r\frac{\rd z_i}{2\pi \i z_i}\;c_\beta(z;q)\;\Delta_T(z)\prod_{i=1}^r z_i^{\sqrt{\beta}(Q-\gamma_0-\sqrt{\beta}r)}\prod_{i=1}^r\prod_{j=1}^N \frac{(u_j y_j z_i^{-1};q)_\infty}{(y_j z_i^{-1};q)_\infty}~,
\ee
where
\be
\Delta_T(z)=\prod_{1\leq i\neq j\leq r}\frac{(z_i z_j^{-1};q)_\infty}{(t z_i z_j^{-1};q)_\infty}
\ee
is the trigonometric Vandermonde-like determinant appearing in the $q$-deformed $\beta$-ensemble studied for example in \cite{Awata:2010yy} in the context of the 5d AGT correspondence. In the special case $t=q^\beta$, $\beta\in\mathbb{Z}_{>0}$, we find
\be
\Delta_E(z)=\prod_{1\leq i\neq j\leq r}\prod_{k=0}^{\beta-1}\Theta(q^k z_j z_i^{-1};q')~,
\ee
which in the trigonometric limit $q'\to 0$ reduces to 
\be
\Delta_T(z)=\prod_{1\leq i\neq j\leq r}\prod_{k=0}^{\beta-1}(1-q^k z_j z_i^{-1})~.
\ee
The latter represents, apart for a factor of $\prod_{i=1}^r z_i^{-\beta(r-1)}$ which can be reabsorbed into the integrand factors, the ordinary $q$-deformation \cite{Mironov:2011dk} of the $\beta$-deformed rational Vandermonde determinant 
\be
\Delta_R(z)=\prod_{1\leq i<j\leq r}(z_i-z_j)^{2\beta}~.
\ee

\section{4d holomorphic blocks}\label{4dholblocks}
\begin{figure}[!ht]
\leavevmode
\begin{center}
\includegraphics[height=0.17\textheight]{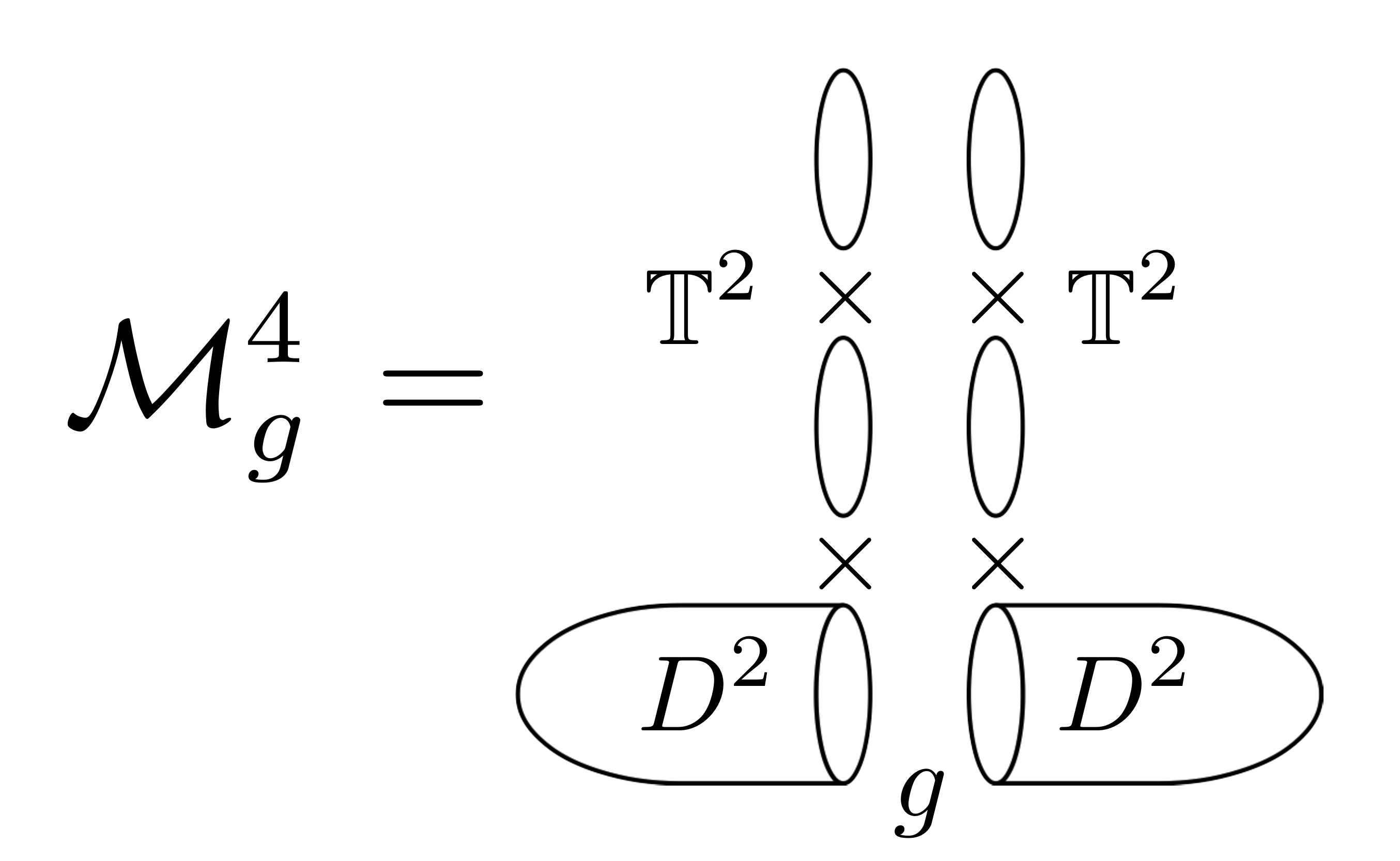}
\end{center}
\caption{Decomposition of $\mathcal{M}^4_g$ into solid tori $D^2\times\mathbb{T}^2$.}
\label{m4g}
\end{figure}

In \cite{Nieri:2015yia} we have analyzed the structure of supersymmetric partition functions of 4d $\mathcal{N}=1$ theories with R-symmetry on compact manifolds ($\mathcal{M}^4_g$). This class of theories can be coupled to new minimal supergravity backgrounds \cite{Dumitrescu:2012ha}, and in the rigid limit \cite{Festuccia:2011ws} 2 supercharges of opposite R-charge can be preserved. In this case $\mathcal{M}^4_g$ must be a Hermitian manifold given by a $\mathbb{T}^2$ fibration over a Riemann surface. When the base has the topology of $S^2$, $\mathcal{M}^4_g$ can be considered to be $S^3\times S^1$, $S^3/\mathbb{Z}_k\times S^1$ or $S^2\times \mathbb{T}^2$. From our viewpoint, $\mathcal{M}^4_g$ is not an elementary geometry in the sense that it admits a Heegaard-like splitting into solid tori $D^2 \times \mathbb{T}^2\simeq \mathbb{R}^2_\epsilon \times \mathbb{T}^2$ (figure \ref{m4g})
\be
\mathcal{M}^4_g\simeq (D^2\times \mathbb{T}^2)\cup_g (D^2\times \mathbb{T}^2)~,
\ee
where $g$ represents a certain element in $SL(3,\mathbb{Z})$ implementing the $\mathbb{T}^3$ boundary homeomorphism realizing the compact geometry and acting on the fibration moduli $\tau,\sigma$. 
In this construction $\tau\propto\epsilon$ is to be identified with the disk equivariant parameter ($\Omega$-deformation) while $\sigma$ with the torus modular parameter. Eventually, one can map $\tau$ and $\sigma$ to the complex structure parameters of $\mathcal{M}^4_g$ (see for example the discussion in \cite{Assel:2014paa,Closset:2013sxa} and references therein).

These geometric observations acquire even more importance if we recall that the compact space partition functions of the class of theories we are considering are quasi-topological objects, as they depend on the complex structure but do not depend on the Hermitian metric \cite{Dumitrescu:2012ha,Closset:2014uda}. Assuming that there are deformations of $\mathcal{M}_g^4$ into a stretched geometry which preserve the complex structure (similarly to the 3d case \cite{Alday:2013lba}), one expects that the associated partition function ($Z$) can be factorized according to the underlying geometric decomposition of $\mathcal{M}_g^4$
\be\label{fact}
Z_{\mathcal{M}_g^4}=\sum_c \Big\| \mathcal{B}_c^{\rm 4d}\Big\|^2_g~,
\ee
where $\mathcal{B}_c^{\rm 4d}$ is identified with the $\mathbb{R}_\epsilon^2\times \mathbb{T}^2$ partition function of the 4d theory and $c$ runs over the supersymmetric vacua of the effective 2d theory. The functions $\mathcal{B}_c^{\rm 4d}$ were called 4d holomorphic blocks \cite{Nieri:2015yia} because the $g$-pairing acts as an involution mapping a block to the conjugate block, where the $g$-action is on $\tau$, $\sigma$ and on a set of variables parametrizing global fugacities.

In \cite{Nieri:2015yia} we have explicitly proved this structure in rank 1 gauge theories and argued its general validity for anomaly free theories by means of other arguments, such as the existence of a commuting set of difference operators annihilating the partition functions, Higgs branch localization \cite{Peelaers:2014ima} or the close relation to $tt^*$ geometries \cite{Cecotti:2013mba}. In fact, for a given gauge theory we have found a specific recipe to compute $\mathcal{B}_c^{\rm 4d}$  through a block integral formalism, similar to that developed in \cite{Beem:2012mb} for the 3d case. The fundamental object of this formalism is an integral kernel $\Upsilon^{\rm 4d}(z)$ whose contour integrals produce the 4d holomorphic blocks\footnote{In order to avoid cluttering, we will denote $\frac{\rd z}{2\pi\i z}\!\!=\!\!\frac{\prod_j \rd z_j}{\prod_j 2\pi \i z_j}$, with the range of $j$ clear from the context.}
\be
\mathcal{B}_c^{\rm 4d}=\oint_{\mathcal{P}_c}\frac{\rd  z}{2\pi\i  z}\Upsilon^{\rm 4d}( z)~,
\ee
where $\mathcal{P}_c$ belongs to a basis of middle dimensional integration cycles in $(\mathbb{C}^{\times})^{|G|}$ which can be determined by the specific matter content and gauge group $G$. The kernel $\Upsilon^{\rm 4d}(z)$ can be assembled using the rules derived in \cite{Nieri:2015yia}, which can be briefly summarized as follows (we refer to \cite{Nieri:2015yia} for a full account):
\begin{itemize}
\item To a vector multiplet we associate a factor of
\be
\mathcal{B}_{\rm vec}^{\rm 4d}( z)=
\prod_{\alpha}\frac{1}{\Gamma(z_\alpha;q_\tau,q_\sigma)}~,
\ee
where $\alpha$ denotes a gauge root. 
\item To a chiral multiplet we associate a factor of
\be
\mathcal{B}_{\rm N}^{\rm 4d}( z, x)=\prod_{\rho}\Gamma(z_\rho x;q_\tau,q_\sigma)~,\quad\textrm{ or }\quad
\mathcal{B}_{\rm D}^{\rm 4d}( z, x)=\prod_{\rho}\frac{1}{\Gamma(q_\tau z_\rho^{-1} x^{-1};q_\tau,q_\sigma)}~,
\ee
where $\rho$ is a gauge weight while $x$ is a global $U(1)$ fugacity.
\end{itemize}
The parameters $q_\tau=\e^{2\pi\i\tau}$ and $q_\sigma=\e^{2\pi\i\sigma}$ can be interpreted as fugacities for rotations on the disk and translations on the torus. We should also observe that the construction of the integral kernel suffers from some ambiguity represented by $q_\tau$-constants.\footnote{We will see that this is also related to the discussion around (\ref{cq}).}

We can now apply the 4d block integral formalism to the higher rank example we have mentioned in the introduction, namely the $U(r)$ theory with $N$ fundamentals and anti-fundamental chirals, 1 adjoint and FI parameter ($\xi$).
Using the rules summarized above, the block integral for this theory can be written as
\be
\mathcal{B}^{\rm 4d}=\oint\prod_{i=1}^r\frac{\rd z_i}{2\pi\i z_i}\Upsilon^{\rm 4d}(z)~,\nn
\ee
\vspace*{-8mm}
\be
\label{blockint}
\ee
\vspace*{-8mm}
\be
\Upsilon^{\rm 4d}(z)=\prod_{1\leq i\neq j\leq r}\frac{\Gamma(t z_i z_j^{-1};q_\tau, q_\sigma)}
{\Gamma(z_i z_j^{-1};q_\tau, q_\sigma)}\prod_{i=1}^{r}z_i^\xi\prod_{a=1}^N
\frac{\Gamma(y_a z_i^{-1};q_\tau,q_\sigma)}{\Gamma(u_a y_a z_i^{-1};q_\tau, q_\sigma)}~,\nn
\ee
where for later convenience the global parameters have been encoded into $t,u, y$.\footnote{The 6d origin of $t$ will be clarified in section \ref{sec6dnek}.}

In this parametrization the 4d block integral (\ref{blockint}) is manifestly equal (up to prefactors) to the correlator (\ref{Npoint}) that we have introduced in the previous section
\be\label{gaugeCFT}
\mathcal{B}^{\rm 4d}\propto 
G^{(N)}_{\gamma_\infty,\gamma_0}~,
\ee
where the identification of parameters is as follows
\be
\renewcommand*{\arraystretch}{1.5}
\begin{array}{c|c|c|c|c|c|c}
\textrm{Gauge theory}~&~ q_\tau~&~q_\sigma~&~t~&~u~&~y~&~\xi\\
\hline
\textrm{EVA}~&~q~&~q'~&~t~&~u~&~y~&~Q-\gamma_0-\sqrt{\beta}r~
\end{array}~~~.
\ee
In particular, the gauge theory integration measure given by the adjoint and vector multiplets and the elliptic Vandermonde-like determinant (\ref{DE}) coming from ``OPE" factor of the screening currents are identified. As we mentioned around (\ref{cq}), the actual measures may differ by $q$-constants, but they give the same result (up to proportionality factors) when integrating along paths enclosing the poles specified in the following (\ref{poles}).\footnote{See also an analogous discussion in appendix C of \cite{Awata:2010yy}.} It then follows that (\ref{blockint}) can be interpreted as the Dotsenko-Fateev representation of the chiral blocks of the EVA, as summarized in (\ref{gaugeCFT}).

We now turn to discussing the integration contour ($\mathcal{C}$), which was left unspecified so far, and the evaluation of the block integral/correlator (\ref{blockint}) by residues. We assume $|q_\tau|<1$, $|q_\sigma|<1$, $|t|<1$ and that they are generic, namely $q_\tau^m\neq q_\sigma^n\neq t^k$ for any $m,n,k\in\mathbb{Z}\backslash\{0\}$. We begin by studying the pole distribution of the block integrand in (\ref{blockint}), focusing on the $u$-independent ones. These are associated to anti-fundamental matter in our conventions, and they determine $\mathcal{C}$ as we are going to explain. The poles coming from the numerator of the matter contribution are located at
\be 
z_i=y_a q_\tau^n q_\sigma^k~,\quad n,k\in\mathbb{Z}_{\geq 0}~.
\ee
Further poles come from the numerator of the integration measure (adjoint) and are determined by the condition
\be
\frac{z_i}{z_j}=t q_\tau^n q_\sigma^k~,\quad n,k\in\mathbb{Z}_{\geq 0}~.
\ee 
Importantly, there are zeros coming from the denominator of the integration measure (vector) whenever
\be
\frac{z_i}{z_j}=q_\tau^n q_\sigma^k~,\quad n,k\in\mathbb{Z}_{\geq 0}~.
\ee
The contour $\mathcal{C}$ is chosen to encircle only the poles of the form
\be
z_i=y_a t ^k q_\tau^n~,\quad n,k\in\mathbb{Z}_{\geq 0}~.
\ee
As explained in detail in \cite{Nieri:2015yia}, this prescription arises by interpreting the contributions from poles containing powers of  $q_\sigma$ as unphysical replicas because of the quasi-periodicity on $\mathbb{T}^2$. Moreover, this prescription works in rank 1 examples \cite{Nieri:2015yia}, and we adopt it also in the present case. In order to completely specify the integration path, we split the integration variables into $N$ groups of $r_{a}$, namely 
\be
r=\sum_{a=1}^{N}r_a~,\quad \{z_i\}=\{z_{(a,\ell)}~|~ a=1,\ldots,N,~\ell=1,\dots,r_a\}~,
\ee 
and we assign a contour to each group. Within each group, the sequence $z_{(a,\ell)}=y_a q_\tau^n$ connects the points $z_{(a,\ell)}=0$ and $z_{(a,\ell)}=y_a$, and the contour is taken to go around that path. In order to understand what are the contributing poles, let us focus on the $a$\textsuperscript{th} group. First of all, we have a permutation symmetry amongst the $z_{(a,\ell)}$ which we fix by starting to integrate from the last variable of the group all the way to the first one, namely we perform the integrations in the order $z_{(a,r_a)},z_{(a,r_a-1)},\ldots,z_{(a,1)}$. For the last variable the contributing poles are just those from the anti-fundamentals, namely $z_{(a,r_a)}=y_a q_\tau^n$. Then we perform the integration over the next-to-last variable $z_{(a,r_a-1)}$. The possible contributing poles arise from the anti-fundamentals at $z_{(a,r_a-1)}=y_a q_\tau^{k}$, or from the the adjoint at $z_{(a,r_a-1)}=y_a t q_\tau^{n+k}$. The first family does not contribute because the condition $\frac{z_{(a,r_a-1)}}{z_{(a,r_a)}}=q_\tau^\mathbb{Z}$ is satisfied and hence the vector contributes with a zero. Similarly, for the variable $z_{(a,r_a-2)}$ we find the contributing poles are those at $z_{(a,r_a-2)}=y_a t^2 q_\tau^{n+k+j}$, and so on. The same reasoning applies to each group, and we can eventually realize that the relevant poles are labeled by an $N$-tuple of Young tableaux $Y^a$ with at most $r_a$ rows of length $Y^a_\ell$
\be\label{poles}
z_{(a,\ell)}=z_{Y^a_\ell}=y_a t^{r_a-\ell}q_\tau^{Y^a_\ell}~,\quad Y^a_\ell\geq Y^a_{\ell+1}~. 
\ee
The sum of the residues of (\ref{blockint}) over these poles can be evaluated by using the properties in (\ref{Gammashift}), and we can finally write
\be\label{vortexsum}
\mathcal{B}^{\rm 4d}_\mathcal{C}={\rm Res}_{z=z_{\vec\emptyset}}\frac{\Upsilon^{\rm 4d}(z)}{z}\times \sum_{\vec Y}\frac{\Upsilon^{\rm 4d}(z)|_{z_{\vec Y}}}{\Upsilon^{\rm 4d}( z)|_{z_{\vec\emptyset}}}~.
\ee
The summands of the series reads as
\be
\frac{\Upsilon^{\rm 4d}(z)|_{z_{\vec Y}}}{\Upsilon^{\rm 4d}(z)|_{z_{\vec\emptyset}}}=q_\tau^{\xi |\vec Y|}\!\!\!\!\!\!\!\prod_{(a,\ell)\neq (b,k)}\!\!\!\frac{\Theta(t y_a y_b^{-1}t^{r_a-r_b-\ell+k};q_\sigma,q_\tau)_{Y^a_\ell-Y^b_k}}{\Theta(y_a y_b^{-1}t^{r_a-r_b-\ell+k};q_\sigma,q_\tau)_{Y^a_\ell-Y^b_k}}
\prod_{a,b,\ell}\!\frac{\Theta(y_b y_a^{-1} t^{-r_a+\ell};q_\sigma,q_\tau)_{-Y^a_\ell}}{\Theta(u_b y_b y_a^{-1} t^{-r_a+\ell};q_\sigma,q_\tau)_{-Y^a_\ell}}~,
\ee
where $|\vec Y|\!=\!\sum_{a=1}^{N}\!\sum_{\ell=1}^{r_a}\!Y^a_\ell$ and the $\Theta$-factorial is defined in (\ref{thetafac}). We see that $\mathcal{B}^{\rm 4d}_\mathcal{C}$ is similar to a multiple elliptic hypergeometric series studied for example in \cite{2003math......3204S}. 

In \cite{Nieri:2015yia} we have shown that the Abelian blocks ($r=1$) are annihilated by a difference operator which is an elliptic deformation (in the shift operator) of the $q$-hypergeometric operator. We believe that there should exist a similar operator annihilating the more general block given in (\ref{vortexsum}), and it would be interesting to determine it. 

The 4d holomorphic block (\ref{vortexsum}) has the form of an elliptic deformation of a vortex partition function \cite{Shadchin:2006yz}, similar to those appearing in \cite{Nieri:2015yia,Peelaers:2014ima,Yoshida:2014qwa,Chen:2014rca}. Given the relation between vortex and instanton counting \cite{Hanany:2003hp,Shadchin:2006yz,Bonelli:2011fq,Fujimori:2015zaa,Dimofte:2010tz}, it is natural to ask whether (\ref{vortexsum}) can be seen as the vortex partition function of a 1/2 BPS codimension 2 theory in $\mathbb{R}^4_{\epsilon_1,\epsilon_2}\times \mathbb{T}^2$. Granted the comment in the previous paragraph, this possibility is also strongly supported by the very well-known fact that partition functions of defect theories obey difference equations \cite{Dimofte:2010tz,Kozcaz:2010af,Bonelli:2011fq,Bonelli:2011wx,Fujimori:2015zaa,Bullimore:2014awa}. Indeed, in the next section we will verify that the elliptic vortex sum in (\ref{vortexsum}) equals the $\mathbb{R}^4\times \mathbb{T}^2$ Nekrasov instanton partition function \cite{Hollowood:2003cv,Hohenegger:2013ala,Haghighat:2013tka} of the $U(N)$ theory with $N$ fundamental and anti-fundamental flavors for particular values of the Coulomb branch parameters of the 6d theory.

\section{6d Nekrasov partition function}\label{sec6dnek}
In this section, we compute the instanton partition function on $\mathbb{R}^4_{\epsilon_1,\epsilon_2}\times \mathbb{T}^2$, which can be defined as the generating function of elliptic genera \cite{Benini:2013nda,Benini:2013xpa} of the instanton moduli space. Our goal is to show that the elliptic Nekrasov instanton partition function reduces to the elliptic vortex partition function (\ref{vortexsum}) at specific points in the Coulomb branch. In fact, in analogy with the lower dimensional cases, these should correspond to the points where vortex solutions exist and where the low energy dynamics can be described by a 1/2 BPS codimension 2 theory on $\mathbb{R}^2_\epsilon\times \mathbb{T}^2$.\footnote{A relation between $\mathcal{N}\!\!=\!\!2$ theories on $S^2\!\times\!\mathbb{T}^2$ and M-strings (see below) is discussed in \cite{Honda:2015yha}.}

\begin{figure}[!ht]
\leavevmode
\begin{center}
\includegraphics[height=0.3\textheight]{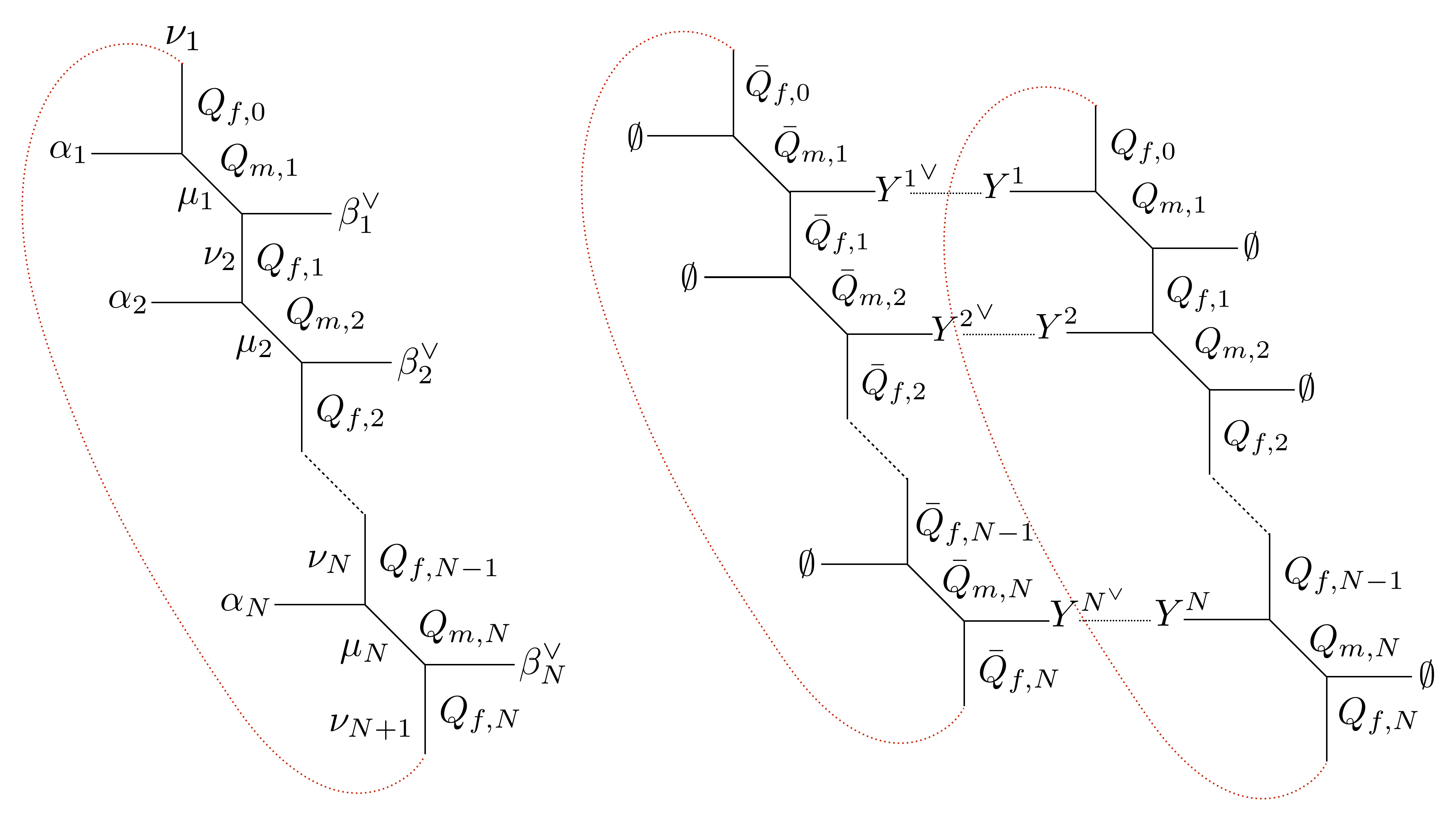}
\end{center}
\caption{Left: The (dual) toric diagram of the periodic strip. Right: Gluing two strips.}
\label{strip}
\end{figure}

There are diverse methods \cite{Hollowood:2003cv} to compute the supersymmetric partition function we are interested in, such as instanton calculus \cite{Nekrasov:2002qd,Nekrasov:2003rj} or topological string methods \cite{Aganagic:2003db,Iqbal:2007ii}. We adopt the second perspective. We start by considering an M-theory setup provided by $M$ parallel M5-branes wrapped on a torus and probing a transverse $A_{N-1}$ singularity as in \cite{Haghighat:2013tka,Hohenegger:2013ala}
\be\label{Mconfig}
\begin{array}{c|cc|cccc|c|cccc}
&\multicolumn{2}{|c|}{\mathbb{T}^2}&\multicolumn{4}{|c|}{\mathbb{R}^4_\parallel}&\mathbb{R}_\perp&\multicolumn{4}{c}{A_{N-1}}\\
\hline
M~{\rm M5}&\bullet&\bullet&\bullet&\bullet&\bullet&\bullet& & & & &
\end{array}~~~.
\ee
Our aim is then to compute the M-theory partition function on such background. The result we are looking for can be found in \cite{Haghighat:2013tka}, here we briefly review the points of that construction which are more relevant for our analysis and adapt to our notation. For coincident M5-branes the configuration leads to a 6d $(1,0)$ superconformal theory. However, one can also consider a deformation away from the superconformal fixed point by separating the M5-branes along a transverse direction, and suspending M2-branes between consecutive M5-branes. The ends of the M2-branes appear as strings from the M5-brane viewpoint, and hence were called ${\rm M}_A$-strings in \cite{Haghighat:2013tka} (the $N>1$ generalization of M-strings \cite{Haghighat:2013gba}). The M-theory partition function can be obtained by computing the BPS degeneracies of the states arising from the M2-branes wrapping $\mathbb{T}^2$. Through a chain of dualities, the M-theory background has a dual description in type IIB string theory as the $(p,q)$-web \cite{Aharony:1997ju,Aharony:1997bh}
\be
\begin{array}{c|cc|cccc|c|ccc}
&S^1&S^1_{(p,q)}&\multicolumn{4}{|c|}{\mathbb{R}^4_\parallel}&\mathbb{R}_{(p,q)}&\multicolumn{3}{c}{\mathbb{R}^3_\perp}\\
\hline
M~{\rm NS5}&\bullet&\bullet&\bullet&\bullet&\bullet&\bullet& & & &\\
N~{\rm D5}&\bullet& &\bullet&\bullet&\bullet&\bullet&\bullet & & &
\end{array}~~~,
\ee
where the sub-index denotes the $(p,q)$-cylinder. The $(p,q)$-web is dual to an elliptically fibered toric Calabi-Yau 3-fold \cite{Leung:1997tw,Kol:1998cf}. This geometry can be used to compute IIA topological string amplitudes by using the refined topological vertex \cite{Awata:2005fa,Iqbal:2007ii,Taki:2007dh}. The basic building block of the geometry is given by the periodic strip depicted in figure \ref{strip} (left), where we have explicitly shown the external Young diagrams and the various K\"ahler parameters. Because of the periodic identification, all the variables of the strip are subject to the equivalence relation $n\sim n+N$ for any sub-index of K\"ahler parameters or Young diagrams. Notice that there is an additional K\"ahler parameter ($Q_{f,0}\sim Q_{f,N}$) and internal Young diagram ($\nu_1\sim \nu_{N+1}$) with respect to the uncompactified strip. By choosing the horizontal direction as the preferred one, the sums over the internal diagrams can be performed through the refined version of the method of \cite{Iqbal:2004ne}, the main difference being the appearance of an extra infinite product taking into account the multi-covering contributions of the basic holomorphic curves due to the periodic identification.
The resulting (normalized) periodic strip amplitude can be written as (\ref{ellstrip})
\begin{multline}
\frac{\mathcal{K}^{\vec\alpha}_{\vec\beta}( Q_m,Q_f;q,t)}{\mathcal{K}^{\vec\emptyset}_{\vec\emptyset}( Q_m,Q_f;q,t)}
=\prod_{a=1}^N\frac{q^{\frac{\|\alpha_a\|^2}{2}}t^{\frac{\|\beta^\vee_a\|^2}{2}}\tilde Z_{\alpha_a}(t,q)\tilde Z_{\beta_a^\vee}(q,t)}{\prod_{k=0}^\infty N_{\beta_a\beta_{a}}(p q'^{k+1};q,t)N_{\alpha_a\alpha_{a}}(q'^{k+1};q,t)}\times\\
\times\frac{\prod_{a,b=1}^N N_{\alpha_{a}\beta_{b}}(p^{\frac{1}{2}}Q_{ab}Q_{m,b};q,t|q')}{\prod_{1\leq a\neq b\leq N}\prod_{k=0}^\infty N_{\beta_a\beta_{b}}(p q'^{k}Q_{ab}Q_{m,a}^{-1}Q_{m,b};q,t)N_{\alpha_a\alpha_{b}}(q'^{k}Q_{ab};q,t)}~,
\end{multline}
with
\be
p=qt^{-1}~,\quad q'=\prod_{k=1}^N Q_{m,k}Q_{f,k}~,\quad Q_{ab}=\left\{\begin{array}{ll}\prod_{k=a}^{b-1}Q_{m,k}Q_{f,k}~,& a<b\\ 
1~,&a=b\\
q' Q_{b a}^{-1}~,& a>b\end{array}\right.~.
\ee
The function $N_{\mu\nu}(Q;q,t|q')$ defined in (\ref{ellnek}) is the elliptic version of the K-theoretic Nekrasov function $N_{\mu\nu}(Q;q,t)$ (\ref{Knek}). The parameters of the $\Omega$-background are identified as $q=\e^{2\pi\i\epsilon_1}$, $t=\e^{-2\pi\i\epsilon_2}$, $q'=\e^{2\pi\i\sigma}$, where $\sigma$ is the elliptic modulus. The 6d theory we are interested in can be engineered by gluing two periodic strips ($M=2$) as in figure \ref{strip} (right) with the constraint
\be\label{2stripconstraint}
Q_{m,a}Q_{f,a}=\bar Q_{m,a+1}\bar Q_{f,a}\quad \Rightarrow\quad \bar Q_{ab}=  Q_{ab}\bar Q_{m,a}\bar Q_{m,b}^{-1}~.
\ee
Setting the external legs to empty Young tableaux, we can compute the $\mathbb{R}^4\times \mathbb{T}^2$ Nekrasov instanton partition function for the $U(N)$ theory with $N$ fundamental and anti-fundamental hypers \cite{Hollowood:2003cv}. The details of the gluing are reported in appendix \ref{appstrip}, the final result is\footnote{For a mathematical definition of the instanton partition function we refer to \cite{Nekrasov:2002qd,Nekrasov:2003rj,2006math.ph...1062O,Nekrasov:2012xe,Nekrasov:2013xda}.}
\be\label{Zinst6d}
\mathcal{Z}^{\mathbb{R}^4\times \mathbb{T}^2}_{\rm inst}=\sum_{\vec Y}\tilde Q_B^{| \vec Y| } 
\prod_{a,b=1}^N\frac{N_{\emptyset Y^b}(A_b^{-1}\bar Q_a;q,t|q') N_{Y^a\emptyset}(A_a Q_b;q,t|q')}{N_{Y^a Y^b}(A_a A_b^{-1};q,t|q')}~,
\ee
where we set
\be
Q_{ab}=\left\{\begin{array}{ll}A_a A_b^{-1}~,& a\leq b\\ 
q' A_a A_b^{-1}~,& a>b\end{array}\right.~,\quad Q_{m,a}=A_a Q_a p^{-\frac{1}{2}}~,\quad \bar Q_{m,a}=A_a^{-1} \bar Q_a p^{-\frac{1}{2}}~.
\ee
Due to $N_{Y^a Y^b}(Q;q,t|0)=N_{Y^a Y^b}(Q;q,t)$, in the decompactification limit $q'\to 0$ we can recognize in the expression above the $\mathbb{R}^4\times S^1$ Nekrasov instanton partition function upon identifying 
\be
A_b=\e^{R a_b}~,\quad  Q_b=\e^{R m_b}~,\quad \bar Q_b=\e^{R \bar m_b }~,\quad \tilde Q_B= \Lambda^{\rm 5d}_{\rm inst}~,
\ee
where $R$ is the scale of the surviving circle, while $a_b$, $m_b$, $\bar m_b$, $\Lambda^{\rm 5d}_{\rm inst}$ are respectively the Coulomb branch parameters, the masses of fundamental and anti-fundamental hypers and the 5d instanton parameter.

We now consider a particular specialization of the parameters $A_a$. If we tune
\be\label{adeg}
A_aQ_a=t^{r_a}~,\quad r_a\in \mathbb{Z}_{>0}~,\quad a=1,\ldots ,N~,
\ee 
the numerator of (\ref{Zinst6d}) yields zero at the box $(r_a+1,1)\in Y^a$, and hence the sum over the Young tableaux is truncated to tableaux with at most $r_a$ rows. When $Y^a$ has at most $r_a$ rows we have
\be
N_{Y^a\emptyset}(Q;q,t|q')=\prod_{i=1}^{r_a}\prod_{j=1}^{Y^a_i}\Theta(Q q^{j-1}t^{1-i};q')=\prod_{i=1}^{r_a}\frac{\Gamma(Qq^{Y^a_i}t^{1-i};q,q')}{\Gamma(Qt^{1-i};q,q')}~,
\ee
where we used $\{\lambda_i-j\}=\{j-1\}$ at fixed $i$ and the definition of the $\Theta$-factorial (\ref{thetafac}). Similarly
\be
N_{\emptyset Y^a}(Q;q,t|q')=\prod_{i=1}^{r_a}\prod_{j=1}^{Y^a_i}\Theta(Qq^{-j}t^{i};q')=\prod_{i=1}^{r_a}\frac{\Gamma(q q' Q^{-1}q^{Y^a_i}t^{-i};q,q')}{\Gamma(q q' Q^{-1}t^{-i};q,q')}~.
\ee
When the diagrams are both non-empty, we can use the identity (for $|q|<1$) \cite{Awata:2008ed}
\be
N_{\mu\nu}(Q;q,t)=\prod_{i,j\geq 1}\frac{(Qq^{\mu_i-\nu_j}t^{j-i+1};q)_\infty}{(Qt^{j-i+1};q)_\infty}\frac{(Qt^{j-i};q)_\infty}{(Qq^{\mu_i-\nu_j}t^{j-i};q)_\infty}~,
\ee
and the definitions (\ref{ellgamma}), (\ref{ellnek}) to write 
\be\label{Nidentity}
N_{\mu\nu}(Q;q,t|q')=\prod_{i,j\geq 1}\frac{\Gamma(Qt^{j-i+1};q,q')}{\Gamma(Qq^{\mu_i-\nu_j}t^{j-i+1};q,q')}\frac{\Gamma(Qq^{\mu_i-\nu_j}t^{j-i};q,q')}{\Gamma(Qt^{j-i};q,q')}~.
\ee
Therefore, when $Y^a$ and $Y^b$ have at most $r_a$ and $r_b$ rows respectively we have
\begin{multline}
N_{Y^a Y^b}(Q;q,t|q')=\prod_{i=1}^{r_a}\prod_{j=1}^{r_b}\frac{\Gamma(Qt^{j-i+1};q,q')}{\Gamma(Qq^{Y^a_i-Y^b_j}t^{j-i+1};q,q')}\frac{\Gamma(Qq^{Y^a_i-Y^b_j}t^{j-i};q,q')}{\Gamma(Qt^{j-i};q,q')}
\times\\
\times N_{Y^a\emptyset}(Qt^{r_b};q,t|q')N_{\emptyset Y^b}(Qt^{-r_a};q,t|q')~,
\end{multline}
where we have divided the infinite products in four regions, namely $(i,j)\in[1,r_a]\times [1,r_b]$, $(i,j)\in[1,r_a]\times [r_b+1,+\infty]$, $(i,j)\in[r_a+1,+\infty]\times [1,r_b]$, $(i,j)\in[r_a+1,+\infty]\times [r_b+1,+\infty]$. The first region contributes with the first factor, the second and the third regions yield the elliptic Nekrasov functions with an empty tableaux, while the fourth region does not contribute. Finally, the evaluation of (\ref{Zinst}) at $A_a=Q_a^{-1}t^{r_a}$ yields
\be\label{6dinst}
\mathcal{Z}_{\rm inst}^{\mathbb{R}^4\times \mathbb{T}^2}=\sum_{\vec Y}\tilde Q_B^{|\vec Y|}\prod_{a,b=1}^N\frac{\mathcal{Z}^{\rm ad}_{Y^a Y^b}}{\mathcal{Z}^{\rm ad}_{\emptyset\emptyset}}\prod_{a=1}^N\frac{\mathcal{Z}^{\rm f}_{Y^a}\mathcal{Z}^{\bar{\rm f}}_{Y^a}}{\mathcal{Z}^{\rm f}_{\emptyset}\mathcal{Z}^{\bar{\rm f}}_{\emptyset}}~,
\ee
where
\be
\begin{split}\label{6dinstdeg}
\mathcal{Z}^{\rm ad}_{Y^aY^b}&=\prod_{i=1}^{r_a}\prod_{j=1}^{r_b}\frac{\Gamma(t Q_b Q_a^{-1}t^{r_a-r_b+j-i}q^{Y^a_i-Y^b_j};q,q')}{\Gamma(Q_b Q_a^{-1}t^{r_a-r_b+j-i}q^{Y^a_i-Y^b_j};q,q')}~,\\
\mathcal{Z}^{\rm f}_{Y^a}\mathcal{Z}^{\bar{\rm f}}_{Y^a}&=\prod_{b=1}^N\prod_{i=1}^{r_a}\frac{\Gamma(Q_b^{-1}Q_a t^{-r_a+i}q^{-Y^a_i};q,q')}{\Gamma(\bar Q_b Q_a t^{-r_a+i}q^{-Y^a_i};q,q')}~.
\end{split}
\ee
We can now easily identify (\ref{6dinst}) with the elliptic vortex sum in (\ref{vortexsum}) provided the following identifications hold\footnote{This relation was anticipated in \cite{Nieri:PHD}, where a review of factorization of supersymmetric partition functions in various dimensions and for diverse compact spaces, 4d and 5d AGT can also be found.}
\be\label{6d4dmap}
\renewcommand*{\arraystretch}{1.5}
\begin{array}{c|c|c|c|c|c|c}
\mathbb{R}^2\times\mathbb{T}^2/{\rm EVA}~&~ q_\tau~&~q_\sigma~&~t~&~u_a~&~y_a~&~q_\tau^\xi\\
\hline
\mathbb{R}^4\times\mathbb{T}^2~&~q~&~q'~&~t~&~Q_a\bar Q_a~&~Q_a^{-1}~&~\tilde Q_B~
\end{array}~~~.
\ee
Moreover, the specialization of the Coulomb branch parameters/internal momenta encodes the rank $r$ of the 4d gauge group/number of screening currents, and it also determines the choice of the 4d block integral/EVA correlator contour through the breaking pattern $r=\sum_{a=1}^{N}r_a$.

\section{Discussion and outlook}\label{out}
In the special case of the 4-point function ($N=2$) with a single screening current ($r=1$), corresponding to the SQED theory with $N=2$ fundamentals and anti-fundamentals, we have shown in \cite{Nieri:2015yia} that the 4d holomorphic block (proportional to the elliptic series ${}_2E_1$) satisfies a $q$-difference equation representing an elliptic deformation of the equation satisfied by the ${}_2\phi_1$ $q$-hypergeometric. In $q$-Virasoro theories this corresponds to the fact that the 4-point correlator has a degenerate insertion at level 2, analogously to the very well-known case of (undeformed) Virasoro theories. It is tempting to make an analogous statement for elliptic Virasoro theories, interpreting the elliptic $q$-difference equation as a decoupling equation for the insertion of a degenerate operator. This is certainly true from the gauge theory viewpoint, as we have shown that the Abelian block arises upon the specialization (\ref{adeg}) $a_1=-m_1-\epsilon_2$, $a_2=-m_2$ of the elliptic Nekrasov instanton partition function, corresponding in the AGT dictionary to the insertion of a level 2 degenerate external momentum \cite{Doroud:2012xw}. In order to fully understand this aspect, a study of the representation theory of the EVA is required.

The results of this work summarized in the ``triality" (\ref{6d4dmap}) and the above observations strongly suggest that, in the spirit of the AGT correspondence, generic chiral blocks of the EVA are described by elliptic Nekrasov instanton partition functions. We hope that this 6d AGT relation and the EVA can be a useful tool for studying certain 6d supersymmetric theories and their defects. It would also be interesting to study the 4d/6d/EVA ``triality" from the perspective of \cite{Dorey:2011pa,Chen:2011sj,Chen:2012we}.

As we mentioned in the main text, our construction of the EVA can be easily generalized to define an elliptic deformation of the $W_M$ algebra. We expect this extended algebra to be important for studying 4d quiver gauge theories and 6d theories engineered by gluing an arbitrary number of periodic strips. It would be also very interesting to develop a stronger version of the 6d AGT correspondence through the identification of compact space partition functions with non-chiral correlators in QFTs with elliptic Virasoro symmetry, along the lines of \cite{Nieri:2013yra,Nieri:2013vba} for the 5d case. 

The EVA may be also interesting from a purely mathematical viewpoint and applications to elliptic integrable systems.  The DVA was introduced to understand the symmetry algebra behind Macdonald polynomials, in analogy with the relation between Jack polynomials and singular vectors of the Virasoro algebra. It was eventually understood \cite{FHSSY} that the DVA and Macdonald polynomials are naturally related to a more elementary algebra, the trigonometric Ding-Iohara algebra \cite{FHHSY}. Elliptic Macdonald functions can be defined as eigenfunctions of the elliptic Macdonald operator. However, their study is much more complicated than in the trigonometric case (see \cite{Bullimore:2014awa,Razamat:2013qfa} for developments from a gauge theory viewpoint). In \cite{saito} an elliptic Ding-Iohara algebra was introduced\footnote{See also \cite{FHHSY} for another elliptic deformation, and \cite{Koroteev:2015dja} for its application to 3d/5d coupled supersymmetric gauge theories and integrable models.} and its connection to elliptic Macdonald functions was established. In appendix \ref{ding}, we show that the EVA can be realized on a tensor product of two Fock representations of the elliptic Ding-Iohara algebra. It is then natural to ask whether elliptic Macdonald functions can be studied by means of the EVA, perhaps through their correspondence with some kind of singular vectors. This perspective may eventually lead to a neat integral representation of elliptic Macdonald functions, as in \cite{Awata:1995eh,Shiraishi:1995rp} for the trigonometric case. 

Finally, in \cite{2011arXiv1106.4088A} the important role of the trigonometric Ding-Iohara algebra for the 5d AGT relation was extensively discussed. It was conjectured (and proved in the Abelian case) that topological string amplitudes on the strip, the basic building block for the 5d Nekrasov instanton partition function, can be computed as matrix elements of a vertex operator intertwining representations of the trigonometric Ding-Iohara algebra. Given the relation amongst the 6d Nekrasov instanton partition function, the EVA and the elliptic Ding-Iohara algebra found in this work, it would be interesting to understand whether periodic strip amplitudes have a similar interpretation.\footnote{Some of these aspects have been considered in the related work \cite{Iqbal:2015fvd}.}

{\it Comment added.} Clavelli-Shapiro trace technique \cite{Clavelli:1973uk} allows torus correlators in $q$-$W_M$ algebras to be interpreted as sphere correlators in elliptic $W_M$ algebras. The advantage of this perspective is that the latter are usually easier to handle. This relation between trigonometric and elliptic algebras is related to the fiber/base duality in the context of 5d or 6d theories arising from toric Calabi-Yau 3-folds with a periodic direction discussed in this paper. However, this duality does not imply that all the elliptic $W_M$ algebra observables can be recast in terms of $q$-$W_M$ algebra ones. This is for instance the case of elliptic torus correlators, which should be interesting for doubly compactified toric geometries. Moreover, it seems to be a non-trivial fact that the elliptic deformation leads to a well-defined associative algebra. 

\acknowledgments
I would like to thank S. Pasquetti for many illuminating discussions and comments about the draft of this paper. I also thank A. Torrielli, M. Zabzine, J. Minahan, and S. Cremonesi for helpful discussions. I also thank the referees for valuable comments. This research was partially supported by the EPSRC PhD studentship EP/K503186/1, and in part by Vetenskapsr\r{a}det 
under grant \#2014- 5517 and by the grant  Geometry and Physics  from the Knut and Alice Wallenberg foundation.

\appendix
\section{Elliptic functions}\label{appell}
In this appendix, we collect useful definitions and properties of some elliptic functions used in the main text. We refer to \cite{naru} for further details. We start by defining the (infinite) $q$-factorial 
\be
(x;q)_\infty=\e^{-{\rm Li}_2(x;q)}~,\quad {\rm Li}_2(x;q)=\sum_{k\geq 1}\frac{x^k}{k(1-q^k)}~.
\ee
In the region $|q|<1$, it has the compact product representation
\be
(x;q)_\infty=\prod_{k\geq 0}(1-q^k x)~,
\ee
which can be extended to the domain $|q|>1$ through
\be
(x;q)_\infty = \frac{1}{\prod_{k\geq 1}(1-q^{-k} x)}~.
\ee
The Jacobi Theta function that we use is defined by
\be\label{Theta}
\Theta(x;q)=(x;q)_\infty(q x^{-1};q)_\infty=\e^{-\sum_{n\neq 0}\frac{x^n}{n(1-q^n)}}~.
\ee
From this simple expression we can realize that throughout this paper we will have consider exponentials of infinite sums running in both directions. These series can be organized by replacing the expansion parameter $x^{n}$ with $\ell^{|n|} x^{n}$, expanding in $\ell$ and then letting $\ell\to 1$. This is for instance how one can verify on a computer Jacobi's triple product identity 
\be
(x;q)_\infty(qx^{-1};q)_\infty=\frac{1}{(q;q)_\infty}\sum_{n\in\mathbb{Z}}(-1)^n q^{n(n-1)/2}x^n~.
\ee
Useful properties of Theta functions are  ($m\in\mathbb{Z}_{\geq 0}$)
\be\label{Thetashift}
\frac{\Theta(q^m x;q)}{\Theta(x;q)}=(-x q^{(m-1)/2})^{-m},\quad 
\frac{\Theta(q^{-m}x;q)}{\Theta(x;q)}=(-x^{-1}q^{(m+1)/2})^{-m}~.
\ee

The double (infinite) $q$-factorial is defined by 
\be
(x;p,q)_\infty=\prod_{j,k\geq 0}(1-p^j q^k x)~,\quad |p|,|q|<1~,
\ee
and it can be extended to other regions by using the representation
\be
(x;p,q)_\infty=\e^{-{\rm Li}_3(x;p,q)}~,\quad {\rm Li}_3(x;p,q)=\sum_{k\geq 1}\frac{x^k}{k(1-p^k)(1-q^k)}~.
\ee
The elliptic Gamma function is defined by
\be\label{ellgamma}
\Gamma(x;p,q)=\frac{(p q x^{-1};p,q)_\infty}{(x;p,q)_\infty}=\e^{\sum_{k\neq 0}\frac{x^k}{k(1-p^k)(1-q^k)}}=\e^{\sum_{k>0}\frac{(q^{-1/2}p^{-1/2}x)^k}{k(q^{k/2}-q^{-k/2})(p^{k/2}-p^{-k/2})}}~.
\ee
Assuming $|p|, |q|<1$, it has zeros and poles at 
\be
\text{zeros}:~ x=p^{m+1}q^{n+1}~,\quad \text{poles}:~ x=p^{-m}q^{-n}~,\quad m,n\in \mathbb{Z}_{\geq 0}~.
\ee 
Useful properties of the elliptic Gamma function are ($m,n\in\mathbb{Z}_{\geq 0}$)
\begin{align}
{\rm Reflection:}~~&\Gamma(x;p,q)\Gamma(p q x^{-1};p,q)=1~,\\
{\rm Shift:}~~\begin{split}
&\frac{\Gamma(p^m q^n x;p,q)}{\Gamma(x;p,q)}=(-x p^{(m-1)/2}q^{(n-1)/2})^{-m n}\Theta(x;p,q)_n\Theta(x;q,p)_m~,\\
&\frac{\Gamma(p^m q^{-n}  x;p,q)}{\Gamma(x;p,q)}=(-x p^{(m-1)/2}q^{-(n+1)/2})^{m n}\frac{\Theta(x;q,p)_m}{\Theta(p q x^{-1};p,q)_n}~,\label{Gammashift}
\end{split}\\
{\rm Residues:}~~&\text{Res}_{x=y p^m q^n}\frac{\Gamma(y x^{-1};p,q)}{x}=\text{Res}_{x=1}\Gamma(x;p,q)~\frac{(-p q~q^{(n-1)/2}p^{(m-1)/2})^{m n}}{\Theta(p q;p,q)_n\Theta(p q;q,p)_m}~.
\end{align} 
Here we  introduced the $\Theta$-factorial 
\begin{align}
\label{thetafac} \Theta(x;p,q)_n&=\frac{\Gamma(q^n x;p,q)}{\Gamma(x;p,q)}=\prod_{k=0}^{n-1}\Theta(x q^k;p)~,\\
\label{thetafacneg}\Theta(x;p,q)_{-n}&=\Theta(q^{-n}x;p,q)_{n}^{-1}~.
\end{align}
Notice that in the limit $p\to 0$ the $\Theta$-factorial reduces to the $q$-factorial
\be
\Theta(x;0,q)_n=(x;q)_n=\prod_{k=0}^{n-1}(1-q^k x)~,
\ee
because 
\be
\Theta(x;0)=1-x~.
\ee

\section{Free boson tools}\label{freeboson}
The master formula when manipulating with bosonic oscillators is the BCH formula. Given two operators ${\sf A}$, ${\sf B}$ such that $[{\sf A},{\sf B}]=c\cdot \mathds{1}$ for some constant $c\in\mathbb{C}$, the BCH formula reduces to 
\be\label{bch}
\e^{\sf A}\e^{\sf B}=\e^{\sf B}\e^{\sf A}\e^{c}~,\quad \e^{-{\sf A}}{\sf B}\e^{\sf A}={\sf B}-[{\sf A},{\sf B}]~.
\ee
Let us now consider an algebra generated by the operators $\{\a_n, n\in\mathbb{Z}\backslash\{0\} \}$ with the defining relations
\be
[\a_m,\a_n]=c_n\delta_{m+n,0}\cdot\mathds{1}~,\quad c_n\in\mathbb{C}~.
\ee
We can construct the following vertex operators  
\be
\V_i=\; :\e^{{\sf v}_i}:~,\quad {\sf v}_i=\sum_{n\neq 0}v_{i,n}\a_n~,
\ee
where $v_{i,n}$ are generic complex numbers and the normal ordering symbol $:~:$ means all the positive modes are moved to the right of all the negative modes. If we denote by $[~]_\pm$ the positive and negative mode parts, we can then write
\be
\V_i=[\V_i]_-[\V_i]_+~, \quad [\V_i]_\pm=\e^{[{\sf v}_i]_\pm}~.
\ee
In the main text, we have to compute correlators containing expressions such as
\be
\prod_{i=1}^M \V_{i}=[\V_1]_-[\V_1]_+[\V_2]_-[\V_2]_+~\cdots~ [\V_M]_-[\V_M]_+~.
\ee
To this end, we bring all the $[\V_i]_+$'s to the right of all the $[\V_j]_-$'s. The first, $[\V_1]_+$, has to cross all the $[\V_i]_-$ with $i=2,\ldots,M$, while it commutes with all the $[\V_i]_+$'s. In the process it produces $\e^{[[{\sf v}_1]_+,[{\sf v}_2]_-]}\cdots \e^{[[{\sf v}_1]_+,[{\sf v}_M]_-]}$ due to (\ref{bch}). The second, $[V_2]_+$, has to cross all the $[\V_i]_-$ with $i=3,\ldots,M$, and so on. Eventually, we get 
\be
\prod_{i=1}^M \V_{i}=\;:\prod_{i=1}^M \V_i:\times \prod_{1\leq i<j\leq M}\e^{[[{\sf v}_i]_+,[{\sf v}_j]_-]}~.
\ee

\section{Free boson representation of the EVA}\label{appfreeEVA}
In this section we give the proof that the current (\ref{EVAff}) satisfies the defining relation (\ref{EVAalgebra}) of the EVA.\footnote{We will follow the lines of the derivation for the $q$-Virasoro case given e.g. in \cite{Odake:1999un}.}  Let us start by computing the ``OPE"
\begin{align}
\T(z)\T(w)&=\Lambda_+(z)\Lambda_+(w)+\Lambda_+(z)\Lambda_-(w)+\Lambda_-(z)\Lambda_+(w)+\Lambda_-(z)\Lambda_-(w)=\nn\\
&=\;:\Lambda_+(z)\Lambda_+(w):f_{+,+}(w/z)^{-1}+:\Lambda_+(z)\Lambda_-(w):f_{+,-}(w/z)^{-1}+\nn\\
&~~~+:\Lambda_-(z)\Lambda_+(w):f_{-,+}(w/z)^{-1}+:\Lambda_-(z)\Lambda_-(w):f_{-,-}(w/z)^{-1}~.
\end{align}
Using (\ref{fgamma}) we have
\begin{align}
f(w/z)\T(z)\T(w)&=\;:\Lambda_+(z)\Lambda_+(w):+:\Lambda_+(z)\Lambda_-(w):\gamma(p^{1/2}w/z)+\nn\\
&~~~+:\Lambda_-(z)\Lambda_+(w):\gamma(p^{-1/2}w/z)+:\Lambda_-(z)\Lambda_-(w):~,
\end{align}
and similarly for $\T(w)\T(z)f(z/w)$ which is obtained by exchanging $z\leftrightarrow w$. Subtracting the two equalities we get
\begin{multline}
f(w/z)\T(z)\T(w)-\T(w)\T(z)f(z/w)=\; :\Lambda_+(z)\Lambda_-(w):(\gamma(p^{1/2}w/z)-\gamma(p^{-1/2}z/w))+\\
+:\Lambda_-(z)\Lambda_+(w):(\gamma(p^{-1/2}w/z)-\gamma(p^{1/2}z/w))~.
\end{multline}
Now using (\ref{gamma2}) we have ($\kappa=\frac{\Theta(q;q')\Theta(t^{-1};q')}{(q';q')^2_\infty\Theta(p,;q')}$)
\begin{align}
\!\!\!\!\!\!\!\!\! f(w/z)\T(z)\T(w)-\T(w)\T(z)f(z/w)&=\; -\kappa:\Lambda_+(z)\Lambda_-(w):(\delta(p w/z)-\delta(w/z))+\nn\\
&~~~ -\kappa:\Lambda_-(z)\Lambda_+(w):(\delta(w/z)-\delta(p^{-1}w/z))=\nn\\
&=-\kappa:\Lambda_+(z)\Lambda_-(p^{-1}z):\delta(p w/z)+\nn\\
&~~~ +\kappa:\Lambda_-(z)\Lambda_+(p z):\delta(p^{-1}w/z)~,
\end{align}
and we can use (\ref{LL1}) to conclude the proof. This is the final result, but the relation (\ref{gamma2}) needs to be considered more carefully. First of all, one may naively conclude that 
\be
\gamma(x)-\gamma(x^{-1})=0~,
\ee
because 
\be
\gamma(x)=\frac{\Theta(p^{1/2}q^{-1}x;q')\Theta(p^{-1/2}q x;q')}{\Theta(p^{1/2}x;q')\Theta(p^{-1/2}x;q')}=\gamma(x^{-1})~,
\ee
where in the last equality we used the property 
\be\label{id1}
\Theta(x^{-1};q')=\Theta(q' x;q' )=-x^{-1} \Theta(x;q')~.
\ee
This is the elliptic analogue of the identity
\be\label{id2}
\frac{1}{1-x}=-\frac{x^{-1}}{1-x^{-1}}~,
\ee
to which it reduces in the limit $q'\to 0$. However, we have to remember that the difference $\gamma(x)-\gamma(x^{-1})$ is coming from subtracting the operator $\T(w)\T(z)f(z/w)$ from the operator $\T(z)\T(w)f(w/z)$, and hence  non-trivial contact terms, signalled by $\delta$ functions, may arise in the process because a radial ordering prescription is needed. In other words, the difference $\gamma(x)-\gamma(x^{-1})$ must be treated in the sense of hyperfunctions (see e.g. \cite{HF}). All in all, the problem is to find a concrete formula for the hyperfunction $\gamma(x)-\gamma(x^{-1})$. Before considering the elliptic case, it might be useful to recall how $\delta$ function terms arise in the $q$-Virasoro limit (see e.g. appendix of  \cite{Odake:1999un}), in which case the $\gamma(x)$ function is substituted by the $q'\to 0$ limit of the one appearing here
\be
\gamma(x)\stackrel{q'\to 0}{\to}\tilde\gamma(x)=\frac{(1-p^{1/2}q^{-1}x)(1-p^{-1/2}q x)}{(1-p^{1/2}x)(1-p^{-1/2}x)}~.
\ee  First of all, the identity  (\ref{id2}) must now be replaced/modified by
\be
\delta(x)=\frac{1}{1-x}+\frac{x^{-1}}{1-x^{-1}}~,
\ee
which can be easily proved by series expanding the two terms in the r.h.s. separately. Now we focus on the factor $(1-p^{1/2}x)$ in the denominator of $\tilde\gamma(x)$, which will be replaced by 
\be
\frac{1}{1-p^{1/2}x}=\delta(p^{1/2}x)-\frac{p^{-1/2}x^{-1}}{1-p^{-1/2}x^{-1}}~,
\ee
and similarly for the factor $(1-p^{1/2}x^{-1})$ in the denominator of $\tilde\gamma(x^{-1})$
\be
\frac{p^{1/2}x^{-1}}{1-p^{1/2}x^{-1}}=\delta(p^{-1/2}x)-\frac{1}{1-p^{-1/2}x}~.
\ee
Now we can take the difference $\tilde\gamma(x)-\tilde\gamma(x^{-1})$ naively, and only the $\delta$ function terms will survive
\be
\tilde\gamma(x)-\tilde\gamma(x^{-1})=-\frac{(1-q)(1-t^{-1})}{(1-p)}(\delta(p^{1/2}x)-\delta(p^{-1/2}x))~.
\ee
In the elliptic case one can repeat exactly the same steps, the crucial point being to find a $\delta$ function representation involving Theta functions which can be used to replace/modify the identity (\ref{id1}). This representation is
\be\label{delta}
\delta(x)=\frac{(q';q')^2_\infty}{\Theta(x;q')}+\frac{(q';q')^2_\infty x^{-1}}{\Theta(x^{-1};q')}~,
\ee
which is given in \cite{saito} ({\it Lemma} 3.5). Now we can use this representation to replace the factor $\Theta(p^{1/2}x;q')$ in the denominator of $\gamma(x)$ with 
\be
\frac{1}{\Theta(p^{1/2}x;q')}=\frac{\delta(p^{1/2}x)}{(q';q')^2_\infty}-\frac{p^{-1/2}x^{-1}}{\Theta(p^{-1/2}x^{-1};q')}~,
\ee 
and similarly for the factor $\Theta(p^{1/2}x^{-1};q')$ in the denominator of $\gamma(x^{-1})$
\be
\frac{p^{1/2}x^{-1}}{\Theta(p^{1/2}x^{-1};q')}=\frac{\delta(p^{-1/2}x)}{(q';q')^2_\infty}-\frac{1}{\Theta(p^{-1/2}x;q')}~.
\ee 
This prescription defines the hyperfunction $\gamma(x)-\gamma(x^{-1})$ in (\ref{gamma2}), which is not identically zero due to contact terms.

\section{Screening current of the EVA}\label{appS}
In this appendix, we provide the explicit verification of (\ref{TSrel}) using (\ref{EVAff}), (\ref{screening}). We set
\be
\Lambda_\sigma(z)=\; :\e^{\lambda_\sigma(z)}: [\Lambda_\sigma(z)]_0~,\quad \S(w)=\; :\e^{{\sf s}(w)}: [\S(w)]_0~,
\ee
where $[~]_{+,-,0}$ denotes the positive, negative or zero mode part. Then we have
\be
[\Lambda_\sigma(z),\S(w)]=\; :\Lambda_\sigma(z)\S(w):\left(t^\sigma\e^{[[\lambda_\sigma(z)]_+,[{\sf s}(w)]_-]}-\e^{[[{\sf s}(w)]_+,[\lambda_\sigma(z)]_-]}\right)~.
\ee
Since
\begin{align}
[[\lambda_\sigma(z)]_+,[{\sf s}(w)]_-]&=\sigma\sum_{n>0}\frac{(p^{\frac{\sigma}{2}}w z^{-1})^n(t^{\frac{n}{2}}-t^{-\frac{n}{2}})}{n(1-q'^n)}+\sigma\sum_{n>0}\frac{(q' p^{-\frac{\sigma}{2}}z w^{-1})^n(t^{-\frac{n}{2}}-t^{\frac{n}{2}})}{n(1-q'^n)}~,\nn\\
[[{\sf s}(w)]_+,[\lambda_\sigma(z)]_-]&=-\sigma\sum_{n>0}\frac{(p^{-\frac{\sigma}{2}}z w^{-1})^n(t^{\frac{n}{2}}-t^{-\frac{n}{2}})}{n(1-q'^n)}-\sigma\sum_{n>0}\frac{(q' p^{\frac{\sigma}{2}}w z^{-1})^n(t^{-\frac{n}{2}}-t^{\frac{n}{2}})}{n(1-q'^n)}~,
\end{align}
we get
\ben
&&[\Lambda_\sigma(z),\S(w)]=\;:\Lambda_\sigma(z)\S(w):F_\sigma(p^{\frac{\sigma}{2}} t^{\frac{1}{2}}w z^{-1})~,
\een
where
\be
F_\sigma(x)=t^\sigma\frac{\Theta(t^{-1} x;q')^\sigma}{\Theta(x;q')^\sigma}-\frac{\Theta(t x^{-1};q')^\sigma}{\Theta(x^{-1};q')^\sigma}~.
\ee
This function should be interpreted again as a hyperfunction (otherwise it is identically zero). Then
\be
F_+(x)=t\frac{\Theta(t^{-1}x;q')}{(q';q')^2_\infty}\delta(x)~,\quad F_-(x)=t^{-1}\frac{\Theta(x;q')}{(q';q')^2_\infty}\delta(x t^{-1})~,
\ee
where we used the representation (\ref{delta}). We can therefore write 
\be
[\T(z),\S(w)]=t^\frac{1}{2}\frac{\Theta(t^{-1};q')}{(q';q')^2_\infty}\sum_{\sigma}~\sigma:\Lambda_\sigma(q^{\frac{\sigma}{2}}w)\S(w):t^{\frac{\sigma}{2}}\delta(q^{\frac{\sigma}{2}}w z^{-1})~.
\ee
It is now easy to verify that
\be
:\Lambda_{+}(q^{\frac{1}{2}}w)\S(w):t^{\frac{1}{2}}=\frac{{\sf X}(q^{\frac{1}{2}}w)}{w}~,\quad :\Lambda_{-}(q^{-\frac{1}{2}}w)\S(w):t^{-\frac{1}{2}}=\frac{{\sf X}(q^{-\frac{1}{2}}w)}{w}~,
\ee
where 
\be
\frac{{\sf X}(w)}{w}=\; :\e^{-\sum_{n\neq 0}\frac{(q^{n/2}p^{-n}+q^{-n/2}) }{(1+p^{-n})(q^{n/2}-q^{-n/2})(1-q'^{|n|})}(w^{-n}\alpha_n-w^n\beta_n)}:\e^{\sqrt{\beta}\Q}w^{\sqrt{\beta}\P}~,
\ee
so that
\be
[\T(z),\S(w)]=(q^{\frac{1}{2}}-q^{-\frac{1}{2}})t^{\frac{1}{2}}\frac{\Theta(t^{-1};q')}{(q';q')^2_\infty}\frac{\rd}{\rd_q w}\delta(w z^{-1}){\sf X}(w)~.
\ee

\section{Relation with the elliptic Ding-Iohara algebra}\label{ding}
The Ding-Iohara algebra \cite{ding}  is an associative unital (functional) $\mathbb{C}$-algebra endowed with a coproduct $\Delta$ (in fact, a Hopf algebra) and defined by a matrix  of analytic structure functions $g_{i j}(z)$ satisfying the property $g_{i j}(z)=g_{j i}(z^{-1})^{-1}$. It provides a generalization of the Drinfeld realization of quantum affine algebras, giving rise to standard examples for particular choices of the structure matrix. 

By considering a single structure function of elliptic type
\be
g(z)=\frac{\Theta(q z;q')\Theta(t^{-1}z;q')\Theta(p^{-1}z;q')}{\Theta(q^{-1}z;q')\Theta(t z;q')\Theta(p z;q')}~,
\ee
the author of \cite{saito}\footnote{We thank S. Shakirov for pointing out this reference to us.} introduced the elliptic Ding-Iohara algebra $\mathcal{U}(q,t,q')$  generated by the coefficients of the currents
\be
x^\pm(z)=\sum_{n}x^\pm_n z^{-n}~,\quad \psi^\pm(z)=\sum_{n}\psi^\pm_n z^{-n}~, \quad n\in\mathbb{Z}~,
\ee
and the invertible central element $\gamma^{1/2}$, subject to the following defining relations
\be
\begin{split}
[\psi^\pm(z),\psi^\pm(w)]&=0~,\\
\psi^+(z)\psi^-(w)&=\psi^-(w)\psi^+(z)\frac{g(\gamma z w^{-1})}{g(\gamma^{-1}z w^{-1})}~,\\
\psi^\rho(z)x^\sigma(w)&=x^\sigma(w)\psi^\rho(z)g(\gamma^{\rho\sigma\frac{1}{2}}z w^{-1})^\sigma~,\quad (\sigma,\rho)\in\{\pm,\pm\}~,\\
x^\pm(z)x^\pm(w)&=x^\pm(w)x^\pm(z)g(z w^{-1})^{\pm 1}~,\\
[x^+(z),x^-(w)]&=\frac{\Theta(q;q')\Theta(t^{-1};q')}{(q';q')_\infty^2\Theta(p;q')}\left(\delta\left(\gamma \frac{w}{z}\right)\psi^+(\gamma^{\frac{1}{2}}w)-\delta\left(\gamma^{-1}\frac{w}{z}\right)\psi^-(\gamma^{-\frac{1}{2}}w)\right)~.
\end{split}
\ee
The coproduct reads
\be
\begin{split}
\Delta(\gamma^{\pm\frac{1}{2}})&=\gamma^{\pm\frac{1}{2}}\otimes \gamma^{\pm\frac{1}{2}}~,\\
\Delta(\psi^\pm(z))&=\psi^\pm(\gamma_{(2)}^{\pm\frac{1}{2}}z)\otimes \psi^\pm(\gamma_{(1)}^{\mp\frac{1}{2}}z)~,\\
\Delta(x^+(z))&=x^+(z)\otimes \mathds{1}+\psi^-(\gamma_{(1)}^{\frac{1}{2}}z)\otimes x^+(\gamma_{(1)}z)~,\\
\Delta(x^-(z))&=\mathds{1}\otimes x^-(z)+x^-(\gamma_{(2)}z)\otimes\psi^+(\gamma_{(2)}^{\frac{1}{2}}z)~,
\end{split}
\ee
where $\gamma^{\pm\frac{1}{2}}_{(1)}=\gamma^{\pm\frac{1}{2}}\otimes\mathds{1}$, $\gamma^{\pm\frac{1}{2}}_{(2)}=\mathds{1}\otimes\gamma^{\pm\frac{1}{2}}$.

We now show that, using the following level 1\footnote{When $\rho(\gamma^{\pm\frac{1}{2}})=p^{\mp\frac{m}{4}}$, $m$ is called the level of the representation.} Fock representation $\rho_y$ ($y\in\mathbb{C^\times}$) of $\mathcal{U}(q,t,q')$ (\cite{saito}, {\it Theorem} 1.2)
\be\label{fockDH}
\rho_y(\gamma^{\pm\frac{1}{2}})=p^{\mp\frac{1}{4}}~,\quad \rho_y(\psi^\pm(z))=\varphi^\pm(z)~,\quad \rho_y(x^+(z))=y\eta(z)~,\quad \rho_y(x^-(z))=y^{-1}\xi(z)~,
\ee
where\footnote{See appendix \ref{freeboson} for notation.} 
\be
\begin{split}
\varphi^\pm(z)&=[\varphi(z)]_\pm~,\\
\varphi(z)&=\;:\e^{\sum_{n\neq 0}\frac{(1-t^n)(p^{-|n|/2}-p^{|n|/2})p^{-|n|/4}z^{-n}}{n(1-q'^{|n|})}\a_n}\e^{-\sum_{n\neq 0}\frac{(1-t^{-n})(p^{-|n|/2}-p^{|n|/2})p^{|n|/4}q'^{|n|}z^{n}}{n(1-q'^{|n|})}\b_n}:~,\\
\eta(z)&=\; :\e^{-\sum_{n\neq 0}\frac{(1-t^n)z^{-n}}{n(1-q'^{|n|})}\a_n}\e^{-\sum_{n\neq 0}\frac{(1-t^{-n})q'^{|n|}z^{n}}{n(1-q'^{|n|})}\b_n}:~,\\
\xi(z)&=\; :\e^{\sum_{n\neq 0}\frac{(1-t^n)p^{-|n|/2}z^{-n}}{n(1-q'^{|n|})}\a_n}\e^{\sum_{n\neq 0}\frac{(1-t^{-n})p^{|n|/2}q'^{|n|}z^{n}}{n(1-q'^{|n|})}\b_n}:~,\\
[\a_m,\a_n]&=m\frac{(1-q'^{|m|})(1-q^{|m|})}{1-t^{|m|}}\delta_{m+n,0}~,\quad m,n\in\mathbb{Z}\backslash\{0\}~,\\
[\b_m,\b_n]&=m\frac{(1-q'^{|m|})(1-q^{|m|})}{(p q')^{|m|}(1-t^{|m|})}\delta_{m+n,0}~,\quad [\a_m,\b_n]=0~,\quad m,n\in\mathbb{Z}\backslash\{0\}~,
\end{split}
\ee
we can give a representation of the EVA algebra defined in (\ref{EVAalgebra}) through the tensor product representation $\rho_{y_1}\otimes \rho_{y_2}$. Our derivation follows the analogous construction of \cite{FHSSY}, where it is shown how to realize the $q$-$W_M$ algebra of \cite{Shiraishi:1995rp,Awata:1995zk} starting from the trigonometric Ding-Iohara algebra, the $q'\to 0$ limit of $\mathcal{U}(q,t,q')$. To begin with, we define the dressed current
\be
{t}(z)=\alpha^-(z)x^+(z)\alpha^+(z)~,
\ee
where the currents $\alpha^\pm (z)$ are defined by means of the modes of $\psi^\pm(z)$ as follows. In analogy with \cite{FHSSY} and having in mind the Fock representation (\ref{fockDH}), we set
\be
\begin{split}
\psi^\pm(z)&=\psi_0^\pm\e^{\pm\sum_{n>0}\Psi_{\pm n}\gamma^{n/2}z^{\mp n}}\e^{\mp\sum_{n>0}\Psi'_{\pm n}\gamma^{-n/2}z^{\pm n}}~,\quad [\psi^+_0,\psi_0^-]=0~,\\
[\Psi_m,\Psi_n]&=\frac{(1-q^{-m})(1-t^m)(1-p^m)}{m(1-q'^{|m|})}(\gamma^m-\gamma^{-m})\gamma^{-|m|}\delta_{m+n,0}~,\\
 [\Psi'_m,\Psi'_n]&=\frac{q'^{|m|}(1-q^{-m})(1-t^m)(1-p^m)}{m(1-q'^{|m|})}
(\gamma^m-\gamma^{-m})\gamma^{|m|}\delta_{m+n,0}~,
\end{split}
\ee
and take 
\be
\alpha^\pm(z)=\e^{\pm\sum_{n>0}\frac{z^{\mp n}}{\gamma^{n}-\gamma^{-n}}\Psi_{\pm n}}\e^{\pm \sum_{n>0}\frac{z^{\pm n}}{\gamma^{n}-\gamma^{-n}}\Psi'_{\pm n}}~.
\ee
Notice that the action of the coproduct on $\Psi_n$, $\Psi'_n$ reads
\be
\Delta(\Psi_n)=\Psi_n\otimes\gamma^{-|n|}+\mathds{1}\otimes\Psi_n~,\quad \Delta(\Psi'_n)=\Psi'_n\otimes\gamma^{|n|}+\mathds{1}\otimes\Psi'_n~,
\ee
while the $\rho_y$ representation is given by $\rho_y(\psi_0^\pm)=1$ and
\be
\rho_y(\Psi_n)=\frac{(1-t^n)(p^{-|n|/2}-p^{|n|/2})}{|n|(1-q'^{|n|})}\a_n~,\quad \rho_y(\Psi'_n)=\frac{q'^{|n|}(1-t^{-n})(p^{-|n|/2}-p^{|n|/2})}{|n|(1-q'^{|n|})}\b_n~.
\ee
We now consider the 2-fold tensor product representation 
\be
\rho_{y_1,y_2}^{(2)}=\rho_{y_1}\otimes\rho_{y_2}\circ \Delta~,
\ee
and we want to compute 
\be
\rho_{y_1,y_2}^{(2)}({ t}(z))=\sum_{i=1,2}y_i\Lambda_i(z)~.
\ee
Notice that we are constructing a level 2 representation since $\rho^{(2)}_{y_1,y_2}(\gamma^{\pm\frac{1}{2}})=p^{\mp\frac{1}{2}}$. In particular, we have
\be
\begin{split}
\rho^{(2)}_{y_1,y_2}(\Psi_n)&=\sum_{i=1,2}\rho^{(2)}_{y_1,y_2,i}(\Psi_n)~,\quad \rho^{(2)}_{y_1,y_2,i}(\Psi_n)=-\frac{(1-t^n)(1-p^{-|n|})p^{\frac{|n|}{2}(2-i+1)}}{|n|(1-q'^{|n|})}\a_{n,i}~,\\
\rho^{(2)}_{y_1,y_2}(\Psi'_n)&=\sum_{i=1,2}\rho^{(2)}_{y_1,y_2,i}(\Psi'_n)~,\quad \rho^{(2)}_{y_1,y_2,i}(\Psi'_n)=\frac{q'^{|n|}(1-t^{-n})(1-p^{|n|})p^{-\frac{|n|}{2}(2-i+1)}}{|n|(1-q'^{|n|})}\b_{n,i}~,
\end{split}
\ee
where the sub-index $i$ denotes that the operators are in the $i$\textsuperscript{th} tensor component. Therefore
\be
\begin{split}
\rho_{y_1,y_2}^{(2)}(\alpha^+(z))&=\prod_{i=1,2}\lambda^+_{i}(z)~,\quad \lambda^+_i(z)=\e^{\sum_{n>0}\frac{z^{-n}}{p^{-n}-p^{n}}\rho_{y_1,y_2,i}^{(2)}(\Psi_n)}\e^{\sum_{n>0}\frac{z^{n}}{p^{-n}-p^{n}}\rho_{y_1,y_2,i}^{(2)}(\Psi'_{n})}~,\\
\rho_{y_1,y_2}^{(2)}(\alpha^-(z))&=\prod_{i=1,2}\lambda^-_{i}(z)~,\quad \lambda^-_i(z)=\e^{-\sum_{n>0}\frac{z^n}{p^{-n}-p^{n}}\rho^{(2)}_{y_1,y_2,i}(\Psi_{-n})}\e^{-\sum_{n>0}\frac{z^{-n}}{p^{-n}-p^{n}}\rho^{(2)}_{y_1,y_2,i}(\Psi'_{-n})}~.
\end{split}
\ee
We also have
\be
\rho_{y_1,y_2}^{(2)}(x^+(z))=\sum_{i=1,2}y_i\tilde \Lambda_i(z)~,\quad \tilde\Lambda_1(z)=\eta(z)\otimes\mathds{1}~,\quad \tilde\Lambda_2(z)=\varphi^-(p^{-1/4}z)\otimes \eta(p^{-1/2}z)~,
\ee
and hence
\be
\Lambda_i(z)=\prod_{j=1,2}\lambda^-_j(z)\tilde \Lambda_i(z)\prod_{k=1,2}\lambda^+_k(z)~.
\ee
Notice that $:\Lambda_1(z)\Lambda_2(p^{-1}z):\;=\mathds{1}$. Finally, we can verify that (\ref{OPELambda}) is satisfied with $\Lambda_{+}(z)\to \Lambda_{1}(z)$, $\Lambda_{-}(z)\to \Lambda_{2}(z)$, namely
\be
\Lambda_{1,2}(z)\Lambda_{1,2}(w)=\; :\Lambda_{1,2}(z)\Lambda_{1,2}(w):f_{\pm,\pm}(w z^{-1})^{-1}~,
\ee
which proves that the current $\rho_{1,1}^{(2)}({ t}(z))$ gives a representation of the EVA algebra. For completeness, let us work  out explicitly the case of $\Lambda_{1}(z)\Lambda_{1}(w)$, the others can be treated similarly. The first tensor component in the above representation of $\Lambda_{1}(z)\Lambda_{1}(w)$ arises from 
\begin{multline}
\lambda^-_1(z)\eta(z)\lambda_1^+(z)\lambda_1^-(w)\eta(w)\lambda^+_1(w)=\\
=\lambda^-_1(z)[\eta(z)]_-\Big([\eta(z)]_+\lambda_1^+(z)\lambda_1^-(w)[\eta(w)]_-\Big)[\eta(w)]_+\lambda^+_1(w)~,
\end{multline}
where we put in brackets the terms which need to be normal ordered. Let us focus on the part containing the $\a$'s. They give four contributions, which sum up to
\be
\e^{-\sum_{m>0}\left(\frac{w}{z}\right)^m\frac{(1-q^m)(1-t^{-m})}{m(1-q'^m)(1-p^{2m})}\left(1-2p^m+p^{2m}+\frac{(1-p^{-m})^2p^{4m}}{(1-p^{2m})}\right)}~.
\ee
The second tensor component arises from 
\be
\lambda_2^-(z)\lambda_2^+(z)\lambda_2^-(w)\lambda_2^+(w)=\lambda_2^-(z)\Big(\lambda_2^+(z)\lambda_2^-(w)\Big)\lambda_2^+(w)~,
\ee
and the normal ordering function from the part containing the $\a$'s reads
\be
\e^{-\sum_{m>0}\left(\frac{w}{z}\right)^m\frac{(1-q^m)(1-t^{-m})}{m(1-q'^m)(1-p^{2m})}\frac{(1-p^{-m})^2p^{3m}}{(1-p^{2m})}}~.
\ee
Analogous results come from the parts containing the $\b$'s but with the replacements $\frac{w}{z}\to\frac{q' z}{w}$, $(q,t,p)\to(q^{-1},t^{-1},p^{-1})$. Multiplying all these contributions together yields
\be
\e^{-\sum_{m>0}\left(\frac{w}{z}\right)^m\frac{(1-q^m)(1-t^{-m})(1-p^m)}{m(1-q'^m)(1-p^{2m})}}\e^{\sum_{m>0}\left(\frac{p^2q' z}{w}\right)^m\frac{(1-q^{-m})(1-t^{m})(1-p^{-m})}{m(1-q'^m)(1-p^{2m})}}=f_{+,+}(w z^{-1})^{-1}~,
\ee
where $f_{+,+}(x)$ is defined in (\ref{ffunction}).

\section{The refined periodic strip}\label{appstrip}
The amplitude of the periodic strip geometry depicted in figure \ref{strip} (left) reads as\footnote{The diagram $\mu^\vee$ is the transpose of $\mu$, and we also have
$|\mu|=\sum_i\mu_i$, $\|\mu\|^2=\sum_i\mu_i^2$. We refer to \cite{Awata:2008ed} for notations and useful combinatorics of Young diagrams.}
\begin{multline}
\!\!\!\mathcal{K}^{\vec\alpha}_{\vec\beta}(Q_m, Q_f;q,t)\!=\!\sum_{\vec\mu,\vec\nu}\prod_{a=1}^{N}(-Q_{m,a})^{|\mu_a|}\!\prod_{b=1}^{N}(-Q_{f,b})^{|\nu_{b+1}|}\!\prod_{a=1}^N C_{\nu_a\mu_a\alpha_a}(t,q)\!\prod_{b=1}^N C_{\nu_{b+1}^\vee\mu_b^\vee\beta_b^\vee}(q,t)\!=\\
=\prod_{a=1}^N q^{\frac{\|\alpha_a\|^2}{2}}t^{\frac{\|\beta^\vee_a\|^2}{2}}\tilde Z_{\alpha_a}(t,q)\tilde Z_{\beta_a^\vee}(q,t)\sum_{\vec\mu,\vec\nu,\vec\eta,\vec\sigma}\prod_{a=1}^{N}(-Q_{m,a})^{|\mu_a|}\prod_{b=1}^{N}(-Q_{f,b})^{|\nu_{b+1}|}\times\\
\times\prod_{a=1}^N s_{\nu^\vee_a/\eta_a}(p^{-\frac{1}{2}}t^{-\rho}q^{-\alpha_a})s_{\mu_a/\eta_a}(q^{-\rho}t^{-\alpha^\vee_a})\prod_{b=1}^N s_{\nu_{b+1}/\sigma_b}(p^{\frac{1}{2}}t q^{-\rho}t^{-\beta^\vee_b})s_{\mu^\vee_b/\sigma_b}(t^{-\rho}q^{-\beta_b})~,
\end{multline}
where we introduced $p=qt^{-1}$, and
\be
\tilde Z_{\mu}(t,q)=\prod_{(i,j)\in\mu}\frac{1}{1-q^{\mu_i-j}t^{\mu_j^\vee-i+1}}~.
\ee 
We need to evaluate
\be
G(x, y,Q_m, Q_f)=\sum_{\vec X,\vec Y}\prod_{a=1}^{2N}(-\tilde Q_a)^{|X_a|}\prod_{a=1}^{2N}s_{X_a/Y_a}(x_a)s_{X^\vee_a/Y_{a+1}}(y_a)~,
\ee
where we grouped the relevant variables according to
\begin{align}
X_a&=\left\{\begin{array}{ll}\mu_\frac{a+1}{2}~,&a\textrm{ odd }\\ 
\nu_{\frac{a}{2}+1}~,&a\textrm{ even }\end{array}\right.~,\quad Y_a=\left\{\begin{array}{ll}\eta_\frac{a+1}{2}~,&a\textrm{ odd }\\
\sigma_\frac{a}{2}~,&a\textrm{ even }\end{array}\right.~,\\
\tilde Q_a&=\left\{\begin{array}{ll}Q_{m,\frac{a+1}{2}}~,& a\textrm{ odd }\\
Q_{f,\frac{a}{2}}~,& a\textrm{ even }\end{array}\right.~,\\
(x_a,y_a)&=\left\{\begin{array}{ll}(q^{-\rho}t^{-\alpha^\vee_\frac{a+1}{2}},t^{-\rho}q^{-\beta_\frac{a+1}{2}})~,& a \textrm{ odd }\\
(q^{-\rho+\frac{1}{2}}t^{-\beta^\vee_\frac{a}{2}-\frac{1}{2}},t^{-\rho+\frac{1}{2}}q^{-\alpha_{\frac{a}{2}+1}-\frac{1}{2}})~,& a\textrm{ even }\end{array}\right.~.
\end{align}
Using standard identities of Schur polynomials \cite{MD} we get\footnote{See appendix B of \cite{Haghighat:2013gba} and appendix A of \cite{Hohenegger:2013ala} for a proof.}
\be
G(x,y, Q_m, Q_f)=\frac{1}{(q';q')_\infty}\prod_{i,j,k=1}^\infty\prod_{a=1}^{2N}\prod_{\ell=1}^N\frac{1-q'^{k-1}\prod_{s=a}^{a+2\ell-2}\tilde Q_s~x_{a,i}y_{a+2\ell-2,j}}{1-q'^{k-1}\prod_{s=a}^{a+2\ell-1}\tilde Q_s~x_{a,i}y_{a+2\ell-1,j}}~,
\ee
where $q'=\prod_{a=1}^{2N}\tilde Q_a=\prod_{a=1}^{N}Q_{m,a}Q_{f,a}$, and hence
\begin{multline}
\mathcal{K}^{\vec\alpha}_{\vec\beta}( Q_m, Q_f;q,t)=\frac{1}{(q';q')_\infty}\prod_{a=1}^N q^{\frac{\|\alpha_a\|^2}{2}}t^{\frac{\|\beta^\vee_a\|^2}{2}}\tilde Z_{\alpha_a}(t,q)\tilde Z_{\beta_a^\vee}(q,t)\times\\
\times\prod_{i,j,k=1}^\infty\prod_{r,\ell=1}^N \frac{(1-q'^{k-1}Q_{r,r+\ell}Q_{m,r}^{-1}q^{-\alpha_{r+\ell,i}+j-\frac{1}{2}}t^{-\beta^\vee_{r,j}+i-\frac{1}{2}})}{(1-q'^{k-1}Q_{r,r+\ell}Q_{m,r}^{-1}Q_{m,r+\ell}q^{-\beta_{r+\ell,i}+j}t^{-\beta^\vee_{r,j}+i-1})}\times\\
\times\frac{(1-q'^{k-1}Q_{r,r+\ell-1}Q_{m,r+\ell-1}q^{-\beta_{r+\ell-1,i}+j-\frac{1}{2}}t^{-\alpha^\vee_{r,j}+i-\frac{1}{2}})}{(1-q'^{k-1}Q_{r,r+\ell}q^{-\alpha_{r+\ell,i}+j-1}t^{-\alpha^\vee_{r,j}+i})}~,
\end{multline}
where
\be
Q_{a,b}=\left\{\begin{array}{ll}\prod_{k=a}^{b-1}Q_{m,k}Q_{f,k}~,& a<b\\ 
q' Q_{b,a}^{-1}~,& a>b\\
1~,& a=b\\
q'~,& b=a+N\end{array}\right.~.
\ee
Using the K-theoretic Nekrasov function (we use the notation of \cite{Awata:2008ed})
\be\label{Knek}
N_{\mu\nu}(Q;q,t)=\prod_{(i,j)\in\mu}(1-Qq^{\mu_i-j}t^{\nu_j^\vee-i+1})\prod_{(i,j)\in\nu}(1-Qq^{-\nu_i+j-1}t^{-\mu_j^\vee+i})~,
\ee
we also have
\be
\prod_{i,j=1}^\infty\frac{1-Qq^{-\nu_i+j-1} t^{-\mu_j^\vee+i}}{1-Qq^{j-1}t^{i}}=N_{\mu\nu}(Q;q,t)~,
\ee
and
\begin{multline}
\frac{\mathcal{K}^{\vec\alpha}_{\vec\beta}(Q_m,Q_f;q,t)}{\mathcal{K}^{\vec\emptyset}_{\vec\emptyset}(Q_m,Q_f;q,t)}
=\prod_{a=1}^N q^{\frac{\|\alpha_a\|^2}{2}}t^{\frac{\|\beta^\vee_a\|^2}{2}}\tilde Z_{\alpha_a}(t,q)\tilde Z_{\beta_a^\vee}(q,t)\times\\
\times\prod_{r,\ell=1}^N\prod_{k=0}^\infty\frac{N_{\beta_r\alpha_{r+\ell}}(p^{\frac{1}{2}} q'^{k}Q_{r,r+\ell}Q_{m,r}^{-1};q,t)N_{\alpha_r\beta_{r+\ell-1}}(p^{\frac{1}{2}} q'^{k}Q_{r,r+\ell-1}Q_{m,r+\ell-1};q,t)}{N_{\beta_r\beta_{r+\ell}}(p q'^{k}Q_{r,r+\ell}Q_{m,r}^{-1}Q_{m,r+\ell};q,t)N_{\alpha_r\alpha_{r+\ell}}(q'^{k}Q_{r,r+\ell};q,t)}=\\
=\prod_{a=1}^N q^{\frac{\|\alpha_a\|^2}{2}}t^{\frac{\|\beta^\vee_a\|^2}{2}}\tilde Z_{\alpha_a}(t,q)\tilde Z_{\beta_a^\vee}(q,t)\prod_{k=0}^\infty \frac{N_{\beta_a\alpha_{a}}(p^{\frac{1}{2}} q'^{k+1}Q_{m,a}^{-1};q,t)N_{\alpha_a\beta_{a}}(p^{\frac{1}{2}}q'^{k}Q_{m,b};q,t)}{N_{\beta_a\beta_{a}}(p q'^{k+1};q,t)N_{\alpha_a\alpha_{a}}(q'^{k+1};q,t)}\times\\
\times\prod_{1\leq a\neq b\leq N}\prod_{k=0}^\infty\frac{N_{\beta_a\alpha_{b}}(p^{\frac{1}{2}} q'^{k}Q_{ab}Q_{m,a}^{-1};q,t)N_{\alpha_a\beta_{b}}(p^{\frac{1}{2}} q'^{k}Q_{a,b}Q_{m,b};q,t)}{N_{\beta_a\beta_{b}}(p q'^{k}Q_{a,b}Q_{m,a}^{-1}Q_{m,b};q,t)N_{\alpha_a\alpha_{b}}(q'^{k}Q_{ab};q,t)}~,
\end{multline}
where we used $q'=Q_{a,b}Q_{b,a}$ for $a\neq b$ and $Q_{a,a+N}=q'$. Introducing the elliptic version of the Nekrasov function\footnote{Notice the symmetry $N_{\mu\nu}(Q;q,t|q')=N_{\nu\mu}(q' p Q^{-1};q,t|q')$.} \footnote{$N_{\mu\nu}(p^\frac{1}{2}x|q')=x^{\frac{|\mu|+|\nu|}{2}}q^{-\frac{\|\nu\|^2-\|\mu\|^2}{4}}t^{-\frac{\|\mu^\vee\|^2-\|\nu^\vee\|^2}{4}}\vartheta_{\mu\nu}(x;q')$ in the notation of \cite{Hohenegger:2013ala}.}
\begin{align}
N_{\mu\nu}(Q;q,t|q')&=\prod_{(i,j)\in\mu}\Theta(Qq^{\mu_i-j}t^{\nu_j^\vee-i+1};q')\prod_{(i,j)\in\nu}\Theta(Qq^{-\nu_i+j-1}t^{-\mu_j^\vee+i};q')~,\\
&=\prod_{k=0}^\infty N_{\mu\nu}(q'^kQ;q,t)N_{\nu\mu}(pq'^{k+1}Q^{-1};q,t)~,\label{ellnek}
\end{align}
we also have
\begin{multline}
\frac{\mathcal{K}^{\vec\alpha}_{\vec\beta}( Q_m, Q_f;q,t)}{\mathcal{K}^{\vec\emptyset}_{\vec\emptyset}(Q_m, Q_f;q,t)}
=\prod_{a=1}^N \frac{q^{\frac{\|\alpha_a\|^2}{2}}t^{\frac{\|\beta^\vee_a\|^2}{2}}\tilde Z_{\alpha_a}(t,q)\tilde Z_{\beta_a^\vee}(q,t)}{N_{\beta_a\beta_{a}}(p q'^{k+1};q,t)N_{\alpha_a\alpha_{a}}(q'^{k+1};q,t)}\times\\
\times\frac{\prod_{a,b=1}^N N_{\alpha_{a}\beta_{b}}(p^{\frac{1}{2}} Q_{a,b}Q_{m,b};q,t|q')}{\prod_{1\leq a\neq b\leq N}\prod_{k=0}^\infty N_{\beta_a\beta_{b}}(p q'^{k}Q_{a,b}Q_{m,a}^{-1}Q_{f,a};q,t)N_{\alpha_a\alpha_{b}}(q'^{k}Q_{a,b};q,t)}~.\label{ellstrip}
\end{multline}
The denominator cannot be expressed solely in terms of the elliptic Nekrasov functions because of the ``wrong" $q$, $t$ shift. This phenomenon has been related to the anomalous modular transformation property of the open periodic strip \cite{Haghighat:2013gba}.

The $\mathbb{R}^4\times\mathbb{T}^2$ Nekrasov instanton partition function of the $U(N)$ theory with $N$ fundamental and anti-fundamental hypers can be computed by gluing two periodic strips, with the result
\begin{align}\label{Zinst}
\mathcal{Z}_{\rm inst}^{\mathbb{R}^4\times\mathbb{T}^2}&=\sum_{\vec Y}\prod_{a=1}^N(-Q_{B,a})^{|Y^a|}\frac{\mathcal{K}^{\vec Y}_{\vec\emptyset}( Q_m, Q_f;q,t)}{\mathcal{K}^{\vec\emptyset}_{\vec\emptyset}(Q_m,Q_f;q,t)}\frac{\mathcal{K}^{\vec\emptyset}_{\vec Y}({\bar Q}_m,{\bar Q}_f;q,t)}{\mathcal{K}^{\vec\emptyset}_{\vec\emptyset}({\bar Q}_m,{\bar Q}_f;q,t)}=\nn\\
&=\sum_{\vec Y}\prod_{a=1}^N(-Q_{B,a})^{|Y^a|}\prod_{a=1}^N\frac{q^{\frac{\|Y^a\|^2}{2}}t^{\frac{\|{Y^a}^\vee\|^2}{2}}\tilde Z_{Y^a}(t,q)\tilde Z_{{Y^a}^\vee}(q,t)}{\prod_{k=0}^\infty N_{Y^aY^a}(p q'^{k+1};q,t)N_{Y^aY^a}(q'^{k+1};q,t)}\times\nn\\
&~~~\times\frac{\prod_{a,b=1}^N N_{\emptyset Y^b}(p^{\frac{1}{2}}Q_{ab}\bar Q_{m,a};q,t|q') N_{Y^a\emptyset}(p^{\frac{1}{2}}Q_{ab}Q_{m,b};q,t|q')}{\prod_{1\leq a\neq b \leq N} N_{Y^a Y^b}(Q_{ab};q,t|q')}~.
\end{align}
Using
\be
q^{\frac{\|\mu\|^2}{2}}t^{\frac{\|\mu^\vee\|^2}{2}}\tilde Z_\mu(t,q)\tilde Z_{\mu^\vee}(q,t)=\frac{(-p^{-\frac{1}{2}})^{|\mu|}}{N_{\mu\mu}(1;q,t)}~,
\ee
we can recast the second line of (\ref{Zinst}) as
\be
\frac{\prod_{a=1}^N (p^{-\frac{1}{2}}Q_{B,a})^{| Y^a|}}{\prod_{a=1}^N N_{Y^aY^a}(1;q,t|q')}~,
\ee
and hence 
\be
\mathcal{Z}_{\rm inst}^{\mathbb{R}^4\times\mathbb{T}^2}=\sum_{\vec Y}\prod_{a=1}^N (p^{-\frac{1}{2}}Q_{B,a})^{|Y^a|}
\prod_{a,b=1}^N\frac{N_{\emptyset Y^b}(p^{\frac{1}{2}}Q_{ab}\bar Q_{m,a};q,t|q') N_{Y^a\emptyset}(p^{\frac{1}{2}} Q_{ab}Q_{m,b};q,t|q')}{N_{Y^a Y^b}(Q_{ab};q,t|q')}~.
\ee
In order to make contact with the field theory language, it is useful to introduce the parametrization 
\be\label{newparam}
Q_{ab}=\left\{\begin{array}{ll}A_a A_b^{-1}~,& a\leq b\\ 
q' A_aA_b^{-1}~,& a>b\end{array}\right.~,\quad Q_{m,a}=A_aQ_a p^{-\frac{1}{2}}~,\quad \bar Q_{m,a}=A_a^{-1} \bar Q_a p^{-\frac{1}{2}}~,
\ee
so that (\ref{2stripconstraint}) is automatically satisfied. Using the shift property
\be 
N_{\mu\nu}(q' Q|q')=N_{\mu\nu}(Q|q')~(-Q)^{-|\mu|-|\nu|}q^{\frac{1}{2}(|\nu|+\|\nu\|^2+|\mu|-\|\mu\|^2)}t^{-\frac{1}{2}(|\mu|-\|\mu^\vee\|^2+|\nu|+\|\nu^\vee\|^2)},
\ee
we can rewrite
\begin{multline}
\mathcal{Z}_{\rm inst}^{\mathbb{R}^4\times\mathbb{T}^2}=\sum_{\vec Y}\prod_{a=1}^N\left( \frac{p^{-\frac{1}{2}}Q_{B,a}}{\prod_{b=1}^N(p^\frac{1}{2}\bar Q_{m,b})}\right)^{|Y^a|} \frac{\prod_{b\leq a}(p^\frac{1}{2}\bar Q_{m,b})^{|Y^a|}}{\prod_{b<a}(p^\frac{1}{2}Q_{m,b})^{|Y^a|}}\times\\
\times \prod_{a,b=1}^N\frac{N_{\emptyset Y^b}(A_b^{-1}\bar Q_a;q,t|q') N_{Y^a\emptyset}(A_a Q_b;q,t|q')}{N_{Y^a Y^b}(A_a A_b^{-1};q,t|q')}~.
\end{multline}
Generically, for $M$ glued strips labeled by $(c)$ we have \cite{Bao:2011rc}
\be
\prod_a {Q^{(c)}_{B,a}}^{|Y_a^{(c)}|}=(Q_{B,1}^{(c)}p^\frac{1}{2}Q_{m,1}^{(c-1)})^{\sum_a|Y^{(c)}_a|}\frac{\prod_{b<a}(p^\frac{1}{2}Q^{(c)}_{m,b})^{|Y^{(c)}_a|}}{\prod_{b \leq a}(p^\frac{1}{2} Q^{(c-1)}_{m,b})^{|Y^{(c)}_a|}}~,
\ee
so that we can define a renormalized expansion parameter $\tilde Q_B$ and write (\ref{Zinst6d}).

\bibliographystyle{utphys}

\begin{thebibliography}{200}
\addtolength{\parskip}{-1ex}

\bibitem{Alday:2009aq}
L.~F. Alday, D.~Gaiotto, and Y.~Tachikawa, ``{Liouville Correlation Functions
  from Four-dimensional Gauge Theories},''
  \href{http://dx.doi.org/10.1007/s11005-010-0369-5}{{\em Lett. Math. Phys.}
  {\bf 91} (2010)  167--197},
\href{http://arxiv.org/abs/0906.3219}{{\tt arXiv:0906.3219 [hep-th]}}.

\bibitem{Gaiotto:2009we}
D.~Gaiotto, ``{N=2 dualities},''
  \href{http://dx.doi.org/10.1007/JHEP08(2012)034}{{\em JHEP} {\bf 08} (2012)
  034},
\href{http://arxiv.org/abs/0904.2715}{{\tt arXiv:0904.2715 [hep-th]}}.

\bibitem{Wyllard:2009hg}
N.~Wyllard, ``{$A_{N-1}$ conformal Toda field theory correlation functions from
  conformal N = 2 SU(N) quiver gauge theories},''
  \href{http://dx.doi.org/10.1088/1126-6708/2009/11/002}{{\em JHEP} {\bf 11}
  (2009)  002},
\href{http://arxiv.org/abs/0907.2189}{{\tt arXiv:0907.2189 [hep-th]}}.

\bibitem{Nekrasov:2002qd}
N.~A. Nekrasov, ``{Seiberg-Witten prepotential from instanton counting},''
  \href{http://dx.doi.org/10.4310/ATMP.2003.v7.n5.a4}{{\em Adv. Theor. Math.
  Phys.} {\bf 7} (2004)  831--864},
\href{http://arxiv.org/abs/hep-th/0206161}{{\tt arXiv:hep-th/0206161
  [hep-th]}}.

\bibitem{Nekrasov:2003rj}
N.~Nekrasov and A.~Okounkov, ``{Seiberg-Witten theory and random partitions},''
\href{http://arxiv.org/abs/hep-th/0306238}{{\tt arXiv:hep-th/0306238
  [hep-th]}}.

\bibitem{Belavin:1984vu}
A.~A. Belavin, A.~M. Polyakov, and A.~B. Zamolodchikov, ``{Infinite Conformal
  Symmetry in Two-Dimensional Quantum Field Theory},''
\href{http://dx.doi.org/10.1016/0550-3213(84)90052-X}{{\em Nucl. Phys.} {\bf
  B241} (1984)  333--380}.

\bibitem{Alday:2009fs}
L.~F. Alday, D.~Gaiotto, S.~Gukov, Y.~Tachikawa, and H.~Verlinde, ``{Loop and
  surface operators in N=2 gauge theory and Liouville modular geometry},''
  \href{http://dx.doi.org/10.1007/JHEP01(2010)113}{{\em JHEP} {\bf 01} (2010)
  113},
\href{http://arxiv.org/abs/0909.0945}{{\tt arXiv:0909.0945 [hep-th]}}.

\bibitem{Drukker:2009id}
N.~Drukker, J.~Gomis, T.~Okuda, and J.~Teschner, ``{Gauge Theory Loop Operators
  and Liouville Theory},''
  \href{http://dx.doi.org/10.1007/JHEP02(2010)057}{{\em JHEP} {\bf 02} (2010)
  057},
\href{http://arxiv.org/abs/0909.1105}{{\tt arXiv:0909.1105 [hep-th]}}.

\bibitem{Frenkel:2015rda}
E.~Frenkel, S.~Gukov, and J.~Teschner, ``{Surface Operators and Separation of
  Variables},''
\href{http://arxiv.org/abs/1506.07508}{{\tt arXiv:1506.07508 [hep-th]}}.

\bibitem{Coman:2015lna}
I.~Coman, M.~Gabella, and J.~Teschner, ``{Line operators in theories of class
  $\mathcal{S}$, quantized moduli space of flat connections, and Toda field
  theory},'' \href{http://dx.doi.org/10.1007/JHEP10(2015)143}{{\em JHEP} {\bf
  10} (2015)  143},
\href{http://arxiv.org/abs/1505.05898}{{\tt arXiv:1505.05898 [hep-th]}}.

\bibitem{Gomis:2014eya}
J.~Gomis and B.~Le~Floch, ``{M2-brane surface operators and gauge theory
  dualities in Toda},''
\href{http://arxiv.org/abs/1407.1852}{{\tt arXiv:1407.1852 [hep-th]}}.

\bibitem{Hosomichi:2010vh}
K.~Hosomichi, S.~Lee, and J.~Park, ``{AGT on the S-duality Wall},''
  \href{http://dx.doi.org/10.1007/JHEP12(2010)079}{{\em JHEP} {\bf 1012} (2010)
   079},
\href{http://arxiv.org/abs/1009.0340}{{\tt arXiv:1009.0340 [hep-th]}}.

\bibitem{Teschner:2012em}
J.~Teschner and G.~Vartanov, ``{6j symbols for the modular double, quantum
  hyperbolic geometry, and supersymmetric gauge theories},''
  \href{http://dx.doi.org/10.1007/s11005-014-0684-3}{{\em Lett. Math. Phys.}
  {\bf 104} (2014)  527--551},
\href{http://arxiv.org/abs/1202.4698}{{\tt arXiv:1202.4698 [hep-th]}}.

\bibitem{Doroud:2012xw}
N.~Doroud, J.~Gomis, B.~Le~Floch, and S.~Lee, ``{Exact Results in D=2
  Supersymmetric Gauge Theories},''
  \href{http://dx.doi.org/10.1007/JHEP05(2013)093}{{\em JHEP} {\bf 05} (2013)
  093},
\href{http://arxiv.org/abs/1206.2606}{{\tt arXiv:1206.2606 [hep-th]}}.

\bibitem{Teschner:2014oja}
J.~Teschner, ``{Exact results on N=2 supersymmetric gauge theories},''
\href{http://arxiv.org/abs/1412.7145}{{\tt arXiv:1412.7145 [hep-th]}}.

\bibitem{Okuda:2014fja}
T.~Okuda, ``{Line operators in supersymmetric gauge theories and the 2d-4d
  relation},''
\href{http://arxiv.org/abs/1412.7126}{{\tt arXiv:1412.7126 [hep-th]}}.

\bibitem{Gukov:2014gja}
S.~Gukov, ``{Surface Operators},''
\href{http://arxiv.org/abs/1412.7127}{{\tt arXiv:1412.7127 [hep-th]}}.

\bibitem{Nekrasov:2012xe}
N.~Nekrasov and V.~Pestun, ``{Seiberg-Witten geometry of four dimensional N=2
  quiver gauge theories},''
\href{http://arxiv.org/abs/1211.2240}{{\tt arXiv:1211.2240 [hep-th]}}.

\bibitem{Nekrasov:2013xda}
N.~Nekrasov, V.~Pestun, and S.~Shatashvili, ``{Quantum geometry and quiver
  gauge theories},''
\href{http://arxiv.org/abs/1312.6689}{{\tt arXiv:1312.6689 [hep-th]}}.

\bibitem{Blum:1997mm}
J.~D. Blum and K.~A. Intriligator, ``{New phases of string theory and 6D RG
  fixed points via branes at orbifold singularities},''
  \href{http://dx.doi.org/10.1016/S0550-3213(97)00449-5}{{\em Nucl. Phys.} {\bf
  B506} (1997)  199--222},
\href{http://arxiv.org/abs/hep-th/9705044}{{\tt arXiv:hep-th/9705044
  [hep-th]}}.

\bibitem{Awata:2009ur}
H.~Awata and Y.~Yamada, ``{Five-dimensional AGT Conjecture and the Deformed
  Virasoro Algebra},'' \href{http://dx.doi.org/10.1007/JHEP01(2010)125}{{\em
  JHEP} {\bf 01} (2010)  125},
\href{http://arxiv.org/abs/0910.4431}{{\tt arXiv:0910.4431 [hep-th]}}.

\bibitem{Shiraishi:1995rp}
J.~Shiraishi, H.~Kubo, H.~Awata, and S.~Odake, ``{A Quantum deformation of the
  Virasoro algebra and the Macdonald symmetric functions},''
  \href{http://dx.doi.org/10.1007/BF00398297}{{\em Lett. Math. Phys.} {\bf 38}
  (1996)  33--51},
\href{http://arxiv.org/abs/q-alg/9507034}{{\tt arXiv:q-alg/9507034 [q-alg]}}.

\bibitem{Awata:1995zk}
H.~Awata, H.~Kubo, S.~Odake, and J.~Shiraishi, ``{Quantum $W_N$ algebras and
  Macdonald polynomials},'' \href{http://dx.doi.org/10.1007/BF02102595}{{\em
  Commun. Math. Phys.} {\bf 179} (1996)  401--416},
\href{http://arxiv.org/abs/q-alg/9508011}{{\tt arXiv:q-alg/9508011 [q-alg]}}.

\bibitem{Feigin:1995sf}
B.~Feigin and E.~Frenkel, ``{Quantum W-algebras and elliptic algebras},''
  \href{http://dx.doi.org/10.1007/BF02108819}{{\em Commun. Math. Phys.} {\bf
  178} (1996)  653--678},
\href{http://arxiv.org/abs/q-alg/9508009}{{\tt arXiv:q-alg/9508009 [q-alg]}}.

\bibitem{Awata:2010yy}
H.~Awata and Y.~Yamada, ``{Five-dimensional AGT Relation and the Deformed
  beta-ensemble},'' \href{http://dx.doi.org/10.1143/PTP.124.227}{{\em Prog.
  Theor. Phys.} {\bf 124} (2010)  227--262},
\href{http://arxiv.org/abs/1004.5122}{{\tt arXiv:1004.5122 [hep-th]}}.

\bibitem{Mironov:2011dk}
A.~Mironov, A.~Morozov, S.~Shakirov, and A.~Smirnov, ``{Proving AGT conjecture
  as HS duality: extension to five dimensions},''
  \href{http://dx.doi.org/10.1016/j.nuclphysb.2011.09.021}{{\em Nucl. Phys.}
  {\bf B855} (2012)  128--151},
\href{http://arxiv.org/abs/1105.0948}{{\tt arXiv:1105.0948 [hep-th]}}.

\bibitem{Carlsson:2013jka}
E.~Carlsson, N.~Nekrasov, and A.~Okounkov, ``{Five dimensional gauge theories
  and vertex operators},'' {\em Moscow Math. J.} {\bf 14} (2014) no.~1, 39--61,
\href{http://arxiv.org/abs/1308.2465}{{\tt arXiv:1308.2465 [math.RT]}}.

\bibitem{Zenkevich:2014lca}
Y.~Zenkevich, ``{Generalized Macdonald polynomials, spectral duality for
  conformal blocks and AGT correspondence in five dimensions},''
  \href{http://dx.doi.org/10.1007/JHEP05(2015)131}{{\em JHEP} {\bf 05} (2015)
  131},
\href{http://arxiv.org/abs/1412.8592}{{\tt arXiv:1412.8592 [hep-th]}}.

\bibitem{Morozov:2015xya}
A.~Morozov and Y.~Zenkevich, ``{Decomposing Nekrasov Decomposition},''
\href{http://arxiv.org/abs/1510.01896}{{\tt arXiv:1510.01896 [hep-th]}}.

\bibitem{Nieri:2013yra}
F.~Nieri, S.~Pasquetti, and F.~Passerini, ``{3d and 5d Gauge Theory Partition
  Functions as $q$-deformed CFT Correlators},''
  \href{http://dx.doi.org/10.1007/s11005-014-0727-9}{{\em Lett. Math. Phys.}
  {\bf 105} (2015) no.~1, 109--148},
\href{http://arxiv.org/abs/1303.2626}{{\tt arXiv:1303.2626 [hep-th]}}.

\bibitem{Nieri:2013vba}
F.~Nieri, S.~Pasquetti, F.~Passerini, and A.~Torrielli, ``{5D partition
  functions, q-Virasoro systems and integrable spin-chains},''
  \href{http://dx.doi.org/10.1007/JHEP12(2014)040}{{\em JHEP} {\bf 12} (2014)
  040},
\href{http://arxiv.org/abs/1312.1294}{{\tt arXiv:1312.1294 [hep-th]}}.

\bibitem{Hosomichi:2012ek}
K.~Hosomichi, R.-K. Seong, and S.~Terashima, ``{Supersymmetric Gauge Theories
  on the Five-Sphere},''
  \href{http://dx.doi.org/10.1016/j.nuclphysb.2012.08.007}{{\em Nucl. Phys.}
  {\bf B865} (2012)  376--396},
\href{http://arxiv.org/abs/1203.0371}{{\tt arXiv:1203.0371 [hep-th]}}.

\bibitem{Kallen:2012cs}
J.~K{\"a}ll{\'e}n and M.~Zabzine, ``{Twisted supersymmetric 5D Yang-Mills
  theory and contact geometry},''
  \href{http://dx.doi.org/10.1007/JHEP05(2012)125}{{\em JHEP} {\bf 05} (2012)
  125},
\href{http://arxiv.org/abs/1202.1956}{{\tt arXiv:1202.1956 [hep-th]}}.

\bibitem{Kallen:2012va}
J.~K{\"a}ll{\'e}n, J.~Qiu, and M.~Zabzine, ``{The perturbative partition
  function of supersymmetric 5D Yang-Mills theory with matter on the
  five-sphere},'' \href{http://dx.doi.org/10.1007/JHEP08(2012)157}{{\em JHEP}
  {\bf 08} (2012)  157},
\href{http://arxiv.org/abs/1206.6008}{{\tt arXiv:1206.6008 [hep-th]}}.

\bibitem{Imamura:2012bm}
Y.~Imamura, ``{Perturbative partition function for squashed $S^5$},''
\href{http://arxiv.org/abs/1210.6308}{{\tt arXiv:1210.6308 [hep-th]}}.

\bibitem{Lockhart:2012vp}
G.~Lockhart and C.~Vafa, ``{Superconformal Partition Functions and
  Non-perturbative Topological Strings},''
\href{http://arxiv.org/abs/1210.5909}{{\tt arXiv:1210.5909 [hep-th]}}.

\bibitem{Kim:2012qf}
H.-C. Kim, J.~Kim, and S.~Kim, ``{Instantons on the 5-sphere and M5-branes},''
\href{http://arxiv.org/abs/1211.0144}{{\tt arXiv:1211.0144 [hep-th]}}.

\bibitem{Kim:2012ava}
H.-C. Kim and S.~Kim, ``{M5-branes from gauge theories on the 5-sphere},''
  \href{http://dx.doi.org/10.1007/JHEP05(2013)144}{{\em JHEP} {\bf 05} (2013)
  144},
\href{http://arxiv.org/abs/1206.6339}{{\tt arXiv:1206.6339 [hep-th]}}.

\bibitem{Minahan:2013jwa}
J.~A. Minahan, A.~Nedelin, and M.~Zabzine, ``{5D super Yang-Mills theory and
  the correspondence to AdS$_7$/CFT$_6$},''
  \href{http://dx.doi.org/10.1088/1751-8113/46/35/355401}{{\em J. Phys.} {\bf
  A46} (2013)  355401},
\href{http://arxiv.org/abs/1304.1016}{{\tt arXiv:1304.1016 [hep-th]}}.

\bibitem{Kim:2012gu}
H.-C. Kim, S.-S. Kim, and K.~Lee, ``{5-dim Superconformal Index with Enhanced
  En Global Symmetry},'' \href{http://dx.doi.org/10.1007/JHEP10(2012)142}{{\em
  JHEP} {\bf 10} (2012)  142},
\href{http://arxiv.org/abs/1206.6781}{{\tt arXiv:1206.6781 [hep-th]}}.

\bibitem{Terashima:2012ra}
S.~Terashima, ``{Supersymmetric gauge theories on $S^4$ x $S^1$},''
  \href{http://dx.doi.org/10.1103/PhysRevD.89.125001}{{\em Phys. Rev.} {\bf
  D89} (2014) no.~12, 125001},
\href{http://arxiv.org/abs/1207.2163}{{\tt arXiv:1207.2163 [hep-th]}}.

\bibitem{Iqbal:2012xm}
A.~Iqbal and C.~Vafa, ``{BPS Degeneracies and Superconformal Index in Diverse
  Dimensions},'' \href{http://dx.doi.org/10.1103/PhysRevD.90.105031}{{\em Phys.
  Rev.} {\bf D90} (2014) no.~10, 105031},
\href{http://arxiv.org/abs/1210.3605}{{\tt arXiv:1210.3605 [hep-th]}}.

\bibitem{Benini:2009gi}
F.~Benini, S.~Benvenuti, and Y.~Tachikawa, ``{Webs of five-branes and N=2
  superconformal field theories},''
  \href{http://dx.doi.org/10.1088/1126-6708/2009/09/052}{{\em JHEP} {\bf 09}
  (2009)  052},
\href{http://arxiv.org/abs/0906.0359}{{\tt arXiv:0906.0359 [hep-th]}}.

\bibitem{Kozcaz:2010af}
C.~Kozcaz, S.~Pasquetti, and N.~Wyllard, ``{A \& B model approaches to surface
  operators and Toda theories},''
  \href{http://dx.doi.org/10.1007/JHEP08(2010)042}{{\em JHEP} {\bf 08} (2010)
  042},
\href{http://arxiv.org/abs/1004.2025}{{\tt arXiv:1004.2025 [hep-th]}}.

\bibitem{Mitev:2014isa}
V.~Mitev and E.~Pomoni, ``{Toda 3-Point Functions From Topological Strings},''
  \href{http://dx.doi.org/10.1007/JHEP06(2015)049}{{\em JHEP} {\bf 06} (2015)
  049},
\href{http://arxiv.org/abs/1409.6313}{{\tt arXiv:1409.6313 [hep-th]}}.

\bibitem{Isachenkov:2014eya}
M.~Isachenkov, V.~Mitev, and E.~Pomoni, ``{Toda 3-Point Functions From
  Topological Strings II},''
\href{http://arxiv.org/abs/1412.3395}{{\tt arXiv:1412.3395 [hep-th]}}.

\bibitem{Aganagic:2013tta}
M.~Aganagic, N.~Haouzi, C.~Kozcaz, and S.~Shakirov, ``{Gauge/Liouville
  Triality},''
\href{http://arxiv.org/abs/1309.1687}{{\tt arXiv:1309.1687 [hep-th]}}.

\bibitem{Aganagic:2014kja}
M.~Aganagic and S.~Shakirov, ``{Gauge/Vortex duality and AGT},''
\href{http://arxiv.org/abs/1412.7132}{{\tt arXiv:1412.7132 [hep-th]}}.

\bibitem{Lukyanov:1994re}
S.~L. Lukyanov and Y.~Pugai, ``{Bosonization of ZF algebras: Direction toward
  deformed Virasoro algebra},'' {\em J. Exp. Theor. Phys.} {\bf 82} (1996)
  1021--1045, \href{http://arxiv.org/abs/hep-th/9412128}{{\tt
  arXiv:hep-th/9412128 [hep-th]}}.
[Zh. Eksp. Teor. Fiz.109,1900(1996)].

\bibitem{Beem:2012mb}
C.~Beem, T.~Dimofte, and S.~Pasquetti, ``{Holomorphic Blocks in Three
  Dimensions},'' \href{http://dx.doi.org/10.1007/JHEP12(2014)177}{{\em JHEP}
  {\bf 1412} (2014)  177},
\href{http://arxiv.org/abs/1211.1986}{{\tt arXiv:1211.1986 [hep-th]}}.

\bibitem{Pasquetti:2011fj}
S.~Pasquetti, ``{Factorisation of N = 2 Theories on the Squashed 3-Sphere},''
  \href{http://dx.doi.org/10.1007/JHEP04(2012)120}{{\em JHEP} {\bf 04} (2012)
  120},
\href{http://arxiv.org/abs/1111.6905}{{\tt arXiv:1111.6905 [hep-th]}}.

\bibitem{Yoshida:2014ssa}
Y.~Yoshida and K.~Sugiyama, ``{Localization of 3d $\mathcal{N}=2$
  Supersymmetric Theories on $S^1 \times D^2$},''
\href{http://arxiv.org/abs/1409.6713}{{\tt arXiv:1409.6713 [hep-th]}}.

\bibitem{Dijkgraaf:2009pc}
R.~Dijkgraaf and C.~Vafa, ``{Toda Theories, Matrix Models, Topological Strings,
  and N=2 Gauge Systems},''
\href{http://arxiv.org/abs/0909.2453}{{\tt arXiv:0909.2453 [hep-th]}}.

\bibitem{Dijkgraaf:2002vw}
R.~Dijkgraaf and C.~Vafa, ``{On geometry and matrix models},''
  \href{http://dx.doi.org/10.1016/S0550-3213(02)00764-2}{{\em Nucl. Phys.} {\bf
  B644} (2002)  21--39},
\href{http://arxiv.org/abs/hep-th/0207106}{{\tt arXiv:hep-th/0207106
  [hep-th]}}.

\bibitem{Dijkgraaf:2002fc}
R.~Dijkgraaf and C.~Vafa, ``{Matrix models, topological strings, and
  supersymmetric gauge theories},''
  \href{http://dx.doi.org/10.1016/S0550-3213(02)00766-6}{{\em Nucl. Phys.} {\bf
  B644} (2002)  3--20},
\href{http://arxiv.org/abs/hep-th/0206255}{{\tt arXiv:hep-th/0206255
  [hep-th]}}.

\bibitem{Aganagic:2014oia}
M.~Aganagic, N.~Haouzi, and S.~Shakirov, ``{$A_n$-Triality},''
\href{http://arxiv.org/abs/1403.3657}{{\tt arXiv:1403.3657 [hep-th]}}.

\bibitem{Seiberg:1997zk}
N.~Seiberg, ``{New theories in six-dimensions and matrix description of M-theory on $T^5$ and $T^5/ Z_2$},''
  \href{http://dx.doi.org/10.1016/S0370-2693(97)00805-8}{{\em Phys. Lett.} {\bf
  B408} (1997)  98--104},
\href{http://arxiv.org/abs/hep-th/9705221}{{\tt arXiv:hep-th/9705221
  [hep-th]}}.

\bibitem{Berkooz:1997cq}
M.~Berkooz, M.~Rozali, and N.~Seiberg, ``{Matrix description of M-theory on
  $T^4$ and $T^5$},''
  \href{http://dx.doi.org/10.1016/S0370-2693(97)00800-9}{{\em Phys. Lett.} {\bf
  B408} (1997)  105--110},
\href{http://arxiv.org/abs/hep-th/9704089}{{\tt arXiv:hep-th/9704089
  [hep-th]}}.

\bibitem{Losev:1997hx}
A.~Losev, G.~W. Moore, and S.~L. Shatashvili, ``{M \& m's},''
  \href{http://dx.doi.org/10.1016/S0550-3213(98)00262-4}{{\em Nucl. Phys.} {\bf
  B522} (1998)  105--124},
\href{http://arxiv.org/abs/hep-th/9707250}{{\tt arXiv:hep-th/9707250
  [hep-th]}}.

\bibitem{Aganagic:2015cta}
M.~Aganagic and N.~Haouzi, ``{ADE Little String Theory on a Riemann Surface
  (and Triality)},''
\href{http://arxiv.org/abs/1506.04183}{{\tt arXiv:1506.04183 [hep-th]}}.

\bibitem{Nieri:2015yia}
F.~Nieri and S.~Pasquetti, ``{Factorisation and holomorphic blocks in 4d},''
\href{http://arxiv.org/abs/1507.00261}{{\tt arXiv:1507.00261 [hep-th]}}.

\bibitem{Hollowood:2003cv}
T.~J. Hollowood, A.~Iqbal, and C.~Vafa, ``{Matrix models, geometric engineering
  and elliptic genera},''
  \href{http://dx.doi.org/10.1088/1126-6708/2008/03/069}{{\em JHEP} {\bf 03}
  (2008)  069},
\href{http://arxiv.org/abs/hep-th/0310272}{{\tt arXiv:hep-th/0310272
  [hep-th]}}.

\bibitem{Haghighat:2013tka}
B.~Haghighat, C.~Kozcaz, G.~Lockhart, and C.~Vafa, ``{Orbifolds of
  M-strings},'' \href{http://dx.doi.org/10.1103/PhysRevD.89.046003}{{\em Phys.
  Rev.} {\bf D89} (2014) no.~4, 046003},
\href{http://arxiv.org/abs/1310.1185}{{\tt arXiv:1310.1185 [hep-th]}}.

\bibitem{Hohenegger:2013ala}
S.~Hohenegger and A.~Iqbal, ``{M-strings, elliptic genera and $\mathcal{N} = 4$
  string amplitudes},'' \href{http://dx.doi.org/10.1002/prop.201300035}{{\em
  Fortsch. Phys.} {\bf 62} (2014)  155--206},
\href{http://arxiv.org/abs/1310.1325}{{\tt arXiv:1310.1325 [hep-th]}}.

\bibitem{Gopakumar:1998ki}
R.~Gopakumar and C.~Vafa, ``{On the gauge theory / geometry correspondence},''
  {\em Adv. Theor. Math. Phys.} {\bf 3} (1999)  1415--1443,
\href{http://arxiv.org/abs/hep-th/9811131}{{\tt arXiv:hep-th/9811131
  [hep-th]}}.

\bibitem{Cachazo:2001jy}
F.~Cachazo, K.~A. Intriligator, and C.~Vafa, ``{A Large N duality via a
  geometric transition},''
  \href{http://dx.doi.org/10.1016/S0550-3213(01)00228-0}{{\em Nucl. Phys.} {\bf
  B603} (2001)  3--41},
\href{http://arxiv.org/abs/hep-th/0103067}{{\tt arXiv:hep-th/0103067
  [hep-th]}}.

\bibitem{Aganagic:2002wv}
M.~Aganagic, A.~Klemm, M.~Marino, and C.~Vafa, ``{Matrix model as a mirror of
  Chern-Simons theory},''
  \href{http://dx.doi.org/10.1088/1126-6708/2004/02/010}{{\em JHEP} {\bf 02}
  (2004)  010},
\href{http://arxiv.org/abs/hep-th/0211098}{{\tt arXiv:hep-th/0211098
  [hep-th]}}.

\bibitem{Aganagic:2011mi}
M.~Aganagic, M.~C.~N. Cheng, R.~Dijkgraaf, D.~Krefl, and C.~Vafa, ``{Quantum
  Geometry of Refined Topological Strings},''
  \href{http://dx.doi.org/10.1007/JHEP11(2012)019}{{\em JHEP} {\bf 11} (2012)
  019},
\href{http://arxiv.org/abs/1105.0630}{{\tt arXiv:1105.0630 [hep-th]}}.

\bibitem{Tan:2013xba}
M.-C. Tan, ``{An M-Theoretic Derivation of a 5d and 6d AGT Correspondence, and
  Relativistic and Elliptized Integrable Systems},''
  \href{http://dx.doi.org/10.1007/JHEP12(2013)031}{{\em JHEP} {\bf 12} (2013)
  031},
\href{http://arxiv.org/abs/1309.4775}{{\tt arXiv:1309.4775 [hep-th]}}.

\bibitem{Heckman:2015bfa}
J.~J. Heckman, D.~R. Morrison, T.~Rudelius, and C.~Vafa, ``{Atomic
  Classification of 6D SCFTs},''
  \href{http://dx.doi.org/10.1002/prop.201500024}{{\em Fortsch. Phys.} {\bf 63}
  (2015)  468--530},
\href{http://arxiv.org/abs/1502.05405}{{\tt arXiv:1502.05405 [hep-th]}}.

\bibitem{Heckman:2013pva}
J.~J. Heckman, D.~R. Morrison, and C.~Vafa, ``{On the Classification of 6D
  SCFTs and Generalized ADE Orbifolds},''
  \href{http://dx.doi.org/10.1007/JHEP06(2015)017,
  10.1007/JHEP05(2014)028}{{\em JHEP} {\bf 05} (2014)  028},
  \href{http://arxiv.org/abs/1312.5746}{{\tt arXiv:1312.5746 [hep-th]}}.
[Erratum: JHEP06,017(2015)].

\bibitem{DelZotto:2015rca}
M.~Del~Zotto, C.~Vafa, and D.~Xie, ``{Geometric Engineering, Mirror Symmetry
  and 6d (1,0) $\to$ 4d, N=2},''
\href{http://arxiv.org/abs/1504.08348}{{\tt arXiv:1504.08348 [hep-th]}}.

\bibitem{Bhardwaj:2015xxa}
L.~Bhardwaj, ``{Classification of 6d N=(1,0) gauge theories},''
\href{http://arxiv.org/abs/1502.06594}{{\tt arXiv:1502.06594 [hep-th]}}.

\bibitem{Hohenegger:2015cba}
S.~Hohenegger, A.~Iqbal, and S.-J. Rey, ``{M-strings, monopole strings, and
  modular forms},'' \href{http://dx.doi.org/10.1103/PhysRevD.92.066005}{{\em
  Phys. Rev.} {\bf D92} (2015) no.~6, 066005},
\href{http://arxiv.org/abs/1503.06983}{{\tt arXiv:1503.06983 [hep-th]}}.

\bibitem{Gadde:2015tra}
A.~Gadde, B.~Haghighat, J.~Kim, S.~Kim, G.~Lockhart, and C.~Vafa, ``{6d String
  Chains},''
\href{http://arxiv.org/abs/1504.04614}{{\tt arXiv:1504.04614 [hep-th]}}.

\bibitem{Haghighat:2015coa}
B.~Haghighat, ``{From strings in 6d to strings in 5d},''
\href{http://arxiv.org/abs/1502.06645}{{\tt arXiv:1502.06645 [hep-th]}}.

\bibitem{Ohmori:2015pua}
K.~Ohmori, H.~Shimizu, Y.~Tachikawa, and K.~Yonekura, ``{6d $\mathcal{N}=(1,0)$
  theories on $T^2$ and class S theories: Part I},''
  \href{http://dx.doi.org/10.1007/JHEP07(2015)014}{{\em JHEP} {\bf 07} (2015)
  014},
\href{http://arxiv.org/abs/1503.06217}{{\tt arXiv:1503.06217 [hep-th]}}.

\bibitem{Ohmori:2015pia}
K.~Ohmori, H.~Shimizu, Y.~Tachikawa, and K.~Yonekura, ``{6d
  $\mathcal{N}{=}(1,0)$ theories on $S^1/T^2$ and class S theories: part II},''
\href{http://arxiv.org/abs/1508.00915}{{\tt arXiv:1508.00915 [hep-th]}}.

\bibitem{Zafrir:2015rga}
G.~Zafrir, ``{Brane webs, $5d$ gauge theories and $6d$ $\mathcal{N}$$=(1,0)$
  SCFT's},''
\href{http://arxiv.org/abs/1509.02016}{{\tt arXiv:1509.02016 [hep-th]}}.

\bibitem{Kim:2015fxa}
J.~Kim, S.~Kim, and K.~Lee, ``{Higgsing towards E-strings},''
\href{http://arxiv.org/abs/1510.03128}{{\tt arXiv:1510.03128 [hep-th]}}.

\bibitem{Ohmori:2015tka}
K.~Ohmori and H.~Shimizu, ``{$S^1/T^2$ Compactifications of 6d
  $\mathcal{N}=(1,0)$ Theories and Brane Webs},''
\href{http://arxiv.org/abs/1509.03195}{{\tt arXiv:1509.03195 [hep-th]}}.

\bibitem{Awata:1996fq}
H.~Awata, H.~Kubo, S.~Odake, and J.~Shiraishi, ``{Virasoro-type symmetries in
  solvable models},''
\href{http://arxiv.org/abs/hep-th/9612233}{{\tt arXiv:hep-th/9612233
  [hep-th]}}.

\bibitem{Odake:1999un}
S.~Odake, ``{Beyond CFT: Deformed Virasoro and elliptic algebras},''
  pp.~307--449.
\newblock 1999.
\newblock
\href{http://arxiv.org/abs/hep-th/9910226}{{\tt arXiv:hep-th/9910226
  [hep-th]}}.
\newblock

\bibitem{frenkelresh:1997}
E.~Frenkel and N.~Reshetikhin, ``{Deformations of W-algebras associated to
  simple Lie algebras},''
\href{http://arxiv.org/abs/9708006}{{\tt arXiv:9708006 [q-alg]}}.

\bibitem{Dumitrescu:2012ha}
T.~T. Dumitrescu, G.~Festuccia, and N.~Seiberg, ``{Exploring Curved
  Superspace},'' \href{http://dx.doi.org/10.1007/JHEP08(2012)141}{{\em JHEP}
  {\bf 08} (2012)  141},
\href{http://arxiv.org/abs/1205.1115}{{\tt arXiv:1205.1115 [hep-th]}}.

\bibitem{Festuccia:2011ws}
G.~Festuccia and N.~Seiberg, ``{Rigid Supersymmetric Theories in Curved
  Superspace},'' \href{http://dx.doi.org/10.1007/JHEP06(2011)114}{{\em JHEP}
  {\bf 06} (2011)  114},
\href{http://arxiv.org/abs/1105.0689}{{\tt arXiv:1105.0689 [hep-th]}}.

\bibitem{Assel:2014paa}
B.~Assel, D.~Cassani, and D.~Martelli, ``{Localization on Hopf surfaces},''
  \href{http://dx.doi.org/10.1007/JHEP08(2014)123}{{\em JHEP} {\bf 08} (2014)
  123},
\href{http://arxiv.org/abs/1405.5144}{{\tt arXiv:1405.5144 [hep-th]}}.

\bibitem{Closset:2013sxa}
C.~Closset and I.~Shamir, ``{The $\mathcal{N}=1$ Chiral Multiplet on $T^2\times
  S^2$ and Supersymmetric Localization},''
  \href{http://dx.doi.org/10.1007/JHEP03(2014)040}{{\em JHEP} {\bf 1403} (2014)
   040},
\href{http://arxiv.org/abs/1311.2430}{{\tt arXiv:1311.2430 [hep-th]}}.

\bibitem{Closset:2014uda}
C.~Closset, T.~T. Dumitrescu, G.~Festuccia, and Z.~Komargodski, ``{From Rigid
  Supersymmetry to Twisted Holomorphic Theories},''
  \href{http://dx.doi.org/10.1103/PhysRevD.90.085006}{{\em Phys. Rev.} {\bf
  D90} (2014) no.~8, 085006},
\href{http://arxiv.org/abs/1407.2598}{{\tt arXiv:1407.2598 [hep-th]}}.

\bibitem{Alday:2013lba}
L.~F. Alday, D.~Martelli, P.~Richmond, and J.~Sparks, ``{Localization on
  Three-Manifolds},'' \href{http://dx.doi.org/10.1007/JHEP10(2013)095}{{\em
  JHEP} {\bf 1310} (2013)  095},
\href{http://arxiv.org/abs/1307.6848}{{\tt arXiv:1307.6848 [hep-th]}}.

\bibitem{Peelaers:2014ima}
W.~Peelaers, ``{Higgs branch localization of $ \mathcal{N} $ = 1 theories on
  S$^{3}$ x S$^{1}$},'' \href{http://dx.doi.org/10.1007/JHEP08(2014)060}{{\em
  JHEP} {\bf 08} (2014)  060},
\href{http://arxiv.org/abs/1403.2711}{{\tt arXiv:1403.2711 [hep-th]}}.

\bibitem{Cecotti:2013mba}
S.~Cecotti, D.~Gaiotto, and C.~Vafa, ``{$tt^*$ geometry in 3 and 4
  dimensions},'' \href{http://dx.doi.org/10.1007/JHEP05(2014)055}{{\em JHEP}
  {\bf 05} (2014)  055},
\href{http://arxiv.org/abs/1312.1008}{{\tt arXiv:1312.1008 [hep-th]}}.

\bibitem{2003math......3204S}
V.~P. {Spiridonov}, ``{Theta hypergeometric series},''{\em ArXiv Mathematics
  e-prints} (Mar., 2003)  , \href{http://arxiv.org/abs/math/0303204}{{\tt
  math/0303204}}.

\bibitem{Shadchin:2006yz}
S.~Shadchin, ``{On F-term contribution to effective action},''
  \href{http://dx.doi.org/10.1088/1126-6708/2007/08/052}{{\em JHEP} {\bf 08}
  (2007)  052},
\href{http://arxiv.org/abs/hep-th/0611278}{{\tt arXiv:hep-th/0611278
  [hep-th]}}.

\bibitem{Yoshida:2014qwa}
Y.~Yoshida, ``{Factorization of 4d N=1 superconformal index},''
\href{http://arxiv.org/abs/1403.0891}{{\tt arXiv:1403.0891 [hep-th]}}.

\bibitem{Chen:2014rca}
H.-Y. Chen and H.-Y. Chen, ``{Heterotic Surface Defects and Dualities from
  2d/4d Indices},'' \href{http://dx.doi.org/10.1007/JHEP10(2014)004}{{\em JHEP}
  {\bf 10} (2014)  004},
\href{http://arxiv.org/abs/1407.4587}{{\tt arXiv:1407.4587 [hep-th]}}.

\bibitem{Hanany:2003hp}
A.~Hanany and D.~Tong, ``{Vortices, instantons and branes},''
  \href{http://dx.doi.org/10.1088/1126-6708/2003/07/037}{{\em JHEP} {\bf 07}
  (2003)  037},
\href{http://arxiv.org/abs/hep-th/0306150}{{\tt arXiv:hep-th/0306150
  [hep-th]}}.

\bibitem{Bonelli:2011fq}
G.~Bonelli, A.~Tanzini, and J.~Zhao, ``{Vertices, Vortices and Interacting
  Surface Operators},'' \href{http://dx.doi.org/10.1007/JHEP06(2012)178}{{\em
  JHEP} {\bf 06} (2012)  178},
\href{http://arxiv.org/abs/1102.0184}{{\tt arXiv:1102.0184 [hep-th]}}.

\bibitem{Fujimori:2015zaa}
T.~Fujimori, T.~Kimura, M.~Nitta, and K.~Ohashi, ``{2d Partition Function in
  Omega-background and Vortex/Instanton Correspondence},''
\href{http://arxiv.org/abs/1509.08630}{{\tt arXiv:1509.08630 [hep-th]}}.

\bibitem{Dimofte:2010tz}
T.~Dimofte, S.~Gukov, and L.~Hollands, ``{Vortex Counting and Lagrangian
  3-manifolds},'' \href{http://dx.doi.org/10.1007/s11005-011-0531-8}{{\em Lett.
  Math. Phys.} {\bf 98} (2011)  225--287},
\href{http://arxiv.org/abs/1006.0977}{{\tt arXiv:1006.0977 [hep-th]}}.

\bibitem{Bonelli:2011wx}
G.~Bonelli, A.~Tanzini, and J.~Zhao, ``{The Liouville side of the Vortex},''
  \href{http://dx.doi.org/10.1007/JHEP09(2011)096}{{\em JHEP} {\bf 09} (2011)
  096},
\href{http://arxiv.org/abs/1107.2787}{{\tt arXiv:1107.2787 [hep-th]}}.

\bibitem{Bullimore:2014awa}
M.~Bullimore, H.-C. Kim, and P.~Koroteev, ``{Defects and Quantum Seiberg-Witten
  Geometry},'' \href{http://dx.doi.org/10.1007/JHEP05(2015)095}{{\em JHEP} {\bf
  05} (2015)  095},
\href{http://arxiv.org/abs/1412.6081}{{\tt arXiv:1412.6081 [hep-th]}}.

\bibitem{Benini:2013nda}
F.~Benini, R.~Eager, K.~Hori, and Y.~Tachikawa, ``{Elliptic genera of
  two-dimensional N=2 gauge theories with rank-one gauge groups},''
  \href{http://dx.doi.org/10.1007/s11005-013-0673-y}{{\em Lett. Math. Phys.}
  {\bf 104} (2014)  465--493},
\href{http://arxiv.org/abs/1305.0533}{{\tt arXiv:1305.0533 [hep-th]}}.

\bibitem{Benini:2013xpa}
F.~Benini, R.~Eager, K.~Hori, and Y.~Tachikawa, ``{Elliptic Genera of 2d
  ${\mathcal{N}}$ = 2 Gauge Theories},''
  \href{http://dx.doi.org/10.1007/s00220-014-2210-y}{{\em Commun. Math. Phys.}
  {\bf 333} (2015) no.~3, 1241--1286},
\href{http://arxiv.org/abs/1308.4896}{{\tt arXiv:1308.4896 [hep-th]}}.

\bibitem{Honda:2015yha}
M.~Honda and Y.~Yoshida, ``{Supersymmetric index on $T^2 \times S^2$ and
  elliptic genus},''
\href{http://arxiv.org/abs/1504.04355}{{\tt arXiv:1504.04355 [hep-th]}}.

\bibitem{Aganagic:2003db}
M.~Aganagic, A.~Klemm, M.~Marino, and C.~Vafa, ``{The Topological vertex},''
  \href{http://dx.doi.org/10.1007/s00220-004-1162-z}{{\em Commun. Math. Phys.}
  {\bf 254} (2005)  425--478},
\href{http://arxiv.org/abs/hep-th/0305132}{{\tt arXiv:hep-th/0305132
  [hep-th]}}.

\bibitem{Iqbal:2007ii}
A.~Iqbal, C.~Kozcaz, and C.~Vafa, ``{The Refined topological vertex},''
  \href{http://dx.doi.org/10.1088/1126-6708/2009/10/069}{{\em JHEP} {\bf 10}
  (2009)  069},
\href{http://arxiv.org/abs/hep-th/0701156}{{\tt arXiv:hep-th/0701156
  [hep-th]}}.

\bibitem{Haghighat:2013gba}
B.~Haghighat, A.~Iqbal, C.~Kozcaz, G.~Lockhart, and C.~Vafa, ``{M-Strings},''
  \href{http://dx.doi.org/10.1007/s00220-014-2139-1}{{\em Commun. Math. Phys.}
  {\bf 334} (2015) no.~2, 779--842},
\href{http://arxiv.org/abs/1305.6322}{{\tt arXiv:1305.6322 [hep-th]}}.

\bibitem{Aharony:1997ju}
O.~Aharony and A.~Hanany, ``{Branes, superpotentials and superconformal fixed
  points},'' \href{http://dx.doi.org/10.1016/S0550-3213(97)00472-0}{{\em Nucl.
  Phys.} {\bf B504} (1997)  239--271},
\href{http://arxiv.org/abs/hep-th/9704170}{{\tt arXiv:hep-th/9704170
  [hep-th]}}.

\bibitem{Aharony:1997bh}
O.~Aharony, A.~Hanany, and B.~Kol, ``{Webs of (p,q) five-branes,
  five-dimensional field theories and grid diagrams},''
  \href{http://dx.doi.org/10.1088/1126-6708/1998/01/002}{{\em JHEP} {\bf 01}
  (1998)  002},
\href{http://arxiv.org/abs/hep-th/9710116}{{\tt arXiv:hep-th/9710116
  [hep-th]}}.

\bibitem{Leung:1997tw}
N.~C. Leung and C.~Vafa, ``{Branes and toric geometry},'' {\em Adv. Theor.
  Math. Phys.} {\bf 2} (1998)  91--118,
\href{http://arxiv.org/abs/hep-th/9711013}{{\tt arXiv:hep-th/9711013
  [hep-th]}}.

\bibitem{Kol:1998cf}
B.~Kol and J.~Rahmfeld, ``{BPS spectrum of five-dimensional field theories,
  (p,q) webs and curve counting},''
  \href{http://dx.doi.org/10.1088/1126-6708/1998/08/006}{{\em JHEP} {\bf 08}
  (1998)  006},
\href{http://arxiv.org/abs/hep-th/9801067}{{\tt arXiv:hep-th/9801067
  [hep-th]}}.

\bibitem{Awata:2005fa}
H.~Awata and H.~Kanno, ``{Instanton counting, Macdonald functions and the
  moduli space of D-branes},''
  \href{http://dx.doi.org/10.1088/1126-6708/2005/05/039}{{\em JHEP} {\bf 05}
  (2005)  039},
\href{http://arxiv.org/abs/hep-th/0502061}{{\tt arXiv:hep-th/0502061
  [hep-th]}}.

\bibitem{Taki:2007dh}
M.~Taki, ``{Refined Topological Vertex and Instanton Counting},''
  \href{http://dx.doi.org/10.1088/1126-6708/2008/03/048}{{\em JHEP} {\bf 03}
  (2008)  048},
\href{http://arxiv.org/abs/0710.1776}{{\tt arXiv:0710.1776 [hep-th]}}.

\bibitem{Iqbal:2004ne}
A.~Iqbal and A.-K. Kashani-Poor, ``{The Vertex on a strip},''
  \href{http://dx.doi.org/10.4310/ATMP.2006.v10.n3.a2}{{\em Adv. Theor. Math.
  Phys.} {\bf 10} (2006)  317--343},
\href{http://arxiv.org/abs/hep-th/0410174}{{\tt arXiv:hep-th/0410174
  [hep-th]}}.

\bibitem{2006math.ph...1062O}
A.~{Okounkov}, ``{Random partitions and instanton counting},''{\em ArXiv
  Mathematical Physics e-prints} (Jan., 2006)  ,
  \href{http://arxiv.org/abs/math-ph/0601062}{{\tt math-ph/0601062}}.

\bibitem{Awata:2008ed}
H.~Awata and H.~Kanno, ``{Refined BPS state counting from Nekrasov's formula
  and Macdonald functions},''
  \href{http://dx.doi.org/10.1142/S0217751X09043006}{{\em Int. J. Mod. Phys.}
  {\bf A24} (2009)  2253--2306},
\href{http://arxiv.org/abs/0805.0191}{{\tt arXiv:0805.0191 [hep-th]}}.

\bibitem{Nieri:PHD}
F.~Nieri, {\em {Integrable structures in supersymmetric gauge theories}}.
\newblock PhD thesis, University of Surrey, Guildford, UK, 2015.

\bibitem{Dorey:2011pa}
N.~Dorey, T.~J. Hollowood, and S.~Lee, ``{Quantization of Integrable Systems
  and a 2d/4d Duality},'' \href{http://dx.doi.org/10.1007/JHEP10(2011)077}{{\em
  JHEP} {\bf 10} (2011)  077},
\href{http://arxiv.org/abs/1103.5726}{{\tt arXiv:1103.5726 [hep-th]}}.

\bibitem{Chen:2011sj}
H.-Y. Chen, N.~Dorey, T.~J. Hollowood, and S.~Lee, ``{A New 2d/4d Duality via
  Integrability},'' \href{http://dx.doi.org/10.1007/JHEP09(2011)040}{{\em JHEP}
  {\bf 09} (2011)  040},
\href{http://arxiv.org/abs/1104.3021}{{\tt arXiv:1104.3021 [hep-th]}}.

\bibitem{Chen:2012we}
H.-Y. Chen, T.~J. Hollowood, and P.~Zhao, ``{A 5d/3d duality from relativistic
  integrable system},'' \href{http://dx.doi.org/10.1007/JHEP07(2012)139}{{\em
  JHEP} {\bf 07} (2012)  139},
\href{http://arxiv.org/abs/1205.4230}{{\tt arXiv:1205.4230 [hep-th]}}.

\bibitem{FHSSY}
B.~{Feigin}, A.~{Hoshino}, J.~{Shibahara}, J.~{Shiraishi}, and S.~{Yanagida},
  ``{Kernel function and quantum algebras},''{\em ArXiv e-prints} (Feb., 2010)
  , \href{http://arxiv.org/abs/1002.2485}{{\tt arXiv:1002.2485 [math.QA]}}.

\bibitem{FHHSY}
B.~{Feigin}, K.~{Hashizume}, A.~{Hoshino}, J.~{Shiraishi}, and S.~{Yanagida},
  ``{A commutative algebra on degenerate CP$^{1}$ and Macdonald
  polynomials},''{\em Journal of Mathematical Physics} (Sept., 2009)  ,
  \href{http://arxiv.org/abs/0904.2291}{{\tt arXiv:0904.2291 [math.CO]}}.

\bibitem{Razamat:2013qfa}
S.~S. Razamat, ``{On the $\mathcal{N} =$ 2 superconformal index and
  eigenfunctions of the elliptic RS model},''
  \href{http://dx.doi.org/10.1007/s11005-014-0682-5}{{\em Lett. Math. Phys.}
  {\bf 104} (2014)  673--690},
\href{http://arxiv.org/abs/1309.0278}{{\tt arXiv:1309.0278 [hep-th]}}.

\bibitem{saito}
Y.~{Saito}, ``{Elliptic Ding-Iohara Algebra and the Free Field Realization of
  the Elliptic Macdonald Operator},''{\em ArXiv e-prints} (Jan., 2013)  ,
  \href{http://arxiv.org/abs/1301.4912}{{\tt arXiv:1301.4912 [math.QA]}}.

\bibitem{Koroteev:2015dja}
P.~Koroteev and A.~Sciarappa, ``{Quantum Hydrodynamics from Large-n
  Supersymmetric Gauge Theories},''
\href{http://arxiv.org/abs/1510.00972}{{\tt arXiv:1510.00972 [hep-th]}}.

\bibitem{Awata:1995eh}
H.~Awata, S.~Odake, and J.~Shiraishi, ``{Integral representations of the
  Macdonald symmetric functions},''
  \href{http://dx.doi.org/10.1007/BF02100101}{{\em Commun. Math. Phys.} {\bf
  179} (1996)  647--666},
\href{http://arxiv.org/abs/q-alg/9506006}{{\tt arXiv:q-alg/9506006 [q-alg]}}.

\bibitem{2011arXiv1106.4088A}
H.~{Awata}, B.~{Feigin}, A.~{Hoshino}, M.~{Kanai}, J.~{Shiraishi}, and
  S.~{Yanagida}, ``{Notes on Ding-Iohara algebra and AGT conjecture},''{\em
  ArXiv e-prints} (June, 2011)  , \href{http://arxiv.org/abs/1106.4088}{{\tt
  arXiv:1106.4088 [math-ph]}}.

\bibitem{Iqbal:2015fvd}
A.~Iqbal, C.~Kozcaz, and S.-T. Yau, ``{Elliptic Virasoro Conformal Blocks},''
\href{http://arxiv.org/abs/1511.00458}{{\tt arXiv:1511.00458 [hep-th]}}.

\bibitem{Clavelli:1973uk}
L.~Clavelli and J.~A. Shapiro, ``{Pomeron factorization in general dual
  models},''
\href{http://dx.doi.org/10.1016/0550-3213(73)90113-2}{{\em Nucl. Phys.} {\bf
  B57} (1973)  490--535}.

\bibitem{HF}
U.~Graf, {\em Introduction to Hyperfunctions and Their Integral Transforms}.
\newblock Birkhäuser Basel, 2010.

\bibitem{ding}
J.-t. Ding and K.~Iohara, ``{Generalization and deformation of Drinfeld quantum
  affine algebras},''
\href{http://dx.doi.org/10.1023/A:1007341410987}{{\em Lett. Math. Phys.} {\bf
  41} (1997)  181--193}.

\bibitem{MD}
I.~Macdonald, {\em Symmetric Functions and Hall Polynomials}.
\newblock Oxford mathematical monographs. Clarendon Press, 1998.

\bibitem{Bao:2011rc}
L.~Bao, E.~Pomoni, M.~Taki, and F.~Yagi, ``{M5-Branes, Toric Diagrams and Gauge
  Theory Duality},'' \href{http://dx.doi.org/10.1007/JHEP04(2012)105}{{\em
  JHEP} {\bf 04} (2012)  105},
\href{http://arxiv.org/abs/1112.5228}{{\tt arXiv:1112.5228 [hep-th]}}.

\bibitem{naru}
A.~{Narukawa}, \emph{The modular properties and the integral representations of
  the multiple elliptic gamma functions}, {\emph{ArXiv Mathematics e-prints}
  (June, 2003) }, [\href{http://arxiv.org/abs/math/0306164}{{\tt
  math/0306164}}].

\end{thebibliography}

\providecommand{\href}[2]{#2}\begingroup\raggedright\endgroup

\end{document}